\documentclass[aps,prx,twocolumn,superscriptaddress]{revtex4-1}

\usepackage{marvosym}
\usepackage{amsmath}
\usepackage{amssymb}
\usepackage{natbib}
\usepackage{graphicx}
\usepackage{bm,hyperref}
\usepackage{braket}

\def\tr{\textrm{tr}}
\newcommand{\sgn}{\textrm{sgn}}

\renewcommand{\Re}{\textrm{Re}}
\renewcommand{\Im}{\textrm{Im}}

\renewcommand{\v}[1]{\ensuremath{\mathbf{#1}}}

\usepackage[usenames,dvipsnames]{color}

\def\be{\begin{equation}}

\def\ba{\begin{eqnarray}}
\def\ea{\end{eqnarray}}
\def\rr{{\bm r}}

\def\w{\omega}

\def\e{\epsilon}
\def\ve{\varepsilon}

\def\p{\partial}

\def\jj{\bm{j}}
\def\pp{{\bm p }}
\def\KK{{\bm K }}
\def\rr{{\bm r }}

\def\vv{\bm v}

\def\EE{\bm{E}}

\def\d{\delta}

\def\Tr{\textrm{Tr}}
\def\tr{\textrm{tr}}

\def\bra{\langle}
\def\ket{\rangle}

\begin{document}

\title{Gyrotropic Hall effect in Berry-curved materials}

\author{E. J. K\"onig}
\affiliation{Department of Physics and Astronomy, Rutgers University, Piscataway, New Jersey, 08854, USA }

\author{M. Dzero}
\affiliation{Department of Physics, Kent State University, Kent, OH 44242, USA}

\author{A. Levchenko}
\affiliation{Department of Physics, University of Wisconsin-Madison, Madison, Wisconsin 53706, USA }

\author{D. A. Pesin}
\affiliation{Department of Physics and Astronomy, University of Utah, Salt Lake City, UT 84112, USA}
\affiliation{Department of Physics, University of Virginia, Charlottesville, VA 22904, USA}

\begin{abstract}
We study the ac Hall response induced by passage of dc transport current in two- and three-dimensional metals with gyrotropic point groups -- the gyrotropic Hall effect -- and consider the phenomenon of current-induced optical activity in noncentrosymmetric metals as a physical application of our theory.  While the effect is expected to be present in single crystals of any noncentrosymmetric metal, we expect it to be strongest in enantiomorphic Weyl semimetals. Using the semiclassical kinetic equation approach we present several mechanisms underlying the gyrotropic Hall effect. Amongst them, the intrinsic mechanism is determined by the Berry curvature dipole, while extrinsic impurity-induced processes are related to skew scattering and side jump phenomena. In general, the intrinsic and extrinsic contributions can be of similar magnitude.  We discuss the gyrotropic Hall effect for all frequencies of practical interest, from the DC transport limit, to optical frequencies. We show that for frequencies that are small compared to relevant band splittings, the trace of the gyrotropic Hall tensor in three-dimensional materials is proportional to a topological, quantized Berry charge, and therefore is robust in gyrotropic Weyl systems. This implies that polycrystals of strongly gyrotropic Weyl semimetals will demonstrate strong current-induced optical activity, whereas the response vanishes for polycrystalline ordinary metals. Therefore, the current-induced optical activity can be considered a valuable tool in identifying the topological nature of a material. 
\end{abstract}
\date{\today}
\maketitle

\section{Introduction}\label{sec:intro}

Studies of quantum phenomena emerging from the wave-function geometry of electronic states of matter lie at the forefront of current interest in modern condensed matter physics. The field was established with seminal works on the theory of the integer quantum Hall effect~\cite{TKNN} and its relation~\cite{NiuWu1985} to the geometric phases~\cite{BerryBook}. Further investigations of linear responses related to electronic band geometry have persisted till recent years, where a new set of phenomena associated with band topology of topological insulators as well as gapless topological phases, most notably Weyl semimetals, have become mainstream research subjects (see Refs.~\cite{HasanKaneReview} and~\cite{ArmitageVishwanath2018} for review). 

In recent years, nonlinear optical and transport phenomena have received increasing attention as ones directly related to band geometry~\cite{sipe2000,morimoto2016}. In particular, the nonlinear Hall and photogalvanic effects~\cite{Deyo2009,SodemannFu2015,Orenstein2016, Lindner2017,Morimoto2017CPVE,ma2018,KangMak2018}, and the second harmonic generation~\cite{wu2017,orenstein2018SHG} attracted considerable theoretical and experimental attention in relation to the physics of electronic systems with nontrivial band topology.

Inspired by these advances, in this paper we consider the gyrotropic Hall effect (GHE), i.e. {ac} Hall response in the presence of a background {dc} current in crystals with gyrotropic point groups. This phenomenon is of principal interest  in noncentrosymmetric time reversal invariant materials, where the linear Hall coefficient vanishes by Onsager's relations, hence the GHE is a way to study the band geometry via magneto-optical effects. 

At optical frequencies, the GHE is the same as the phenomenon of current-induced optical activity, which has been studied both theoretically~\cite{baranova1977,Ivchenko1978}, and experimentally~\cite{Ivchenko1979,Farbshtein2012,LeeShan2017}. We reserve a more general name of the GHE to emphasize that it also pertains to the limit of low frequencies, and purely transport measurements. 

Overall, the entirety of nonlinear Hall effects, to which the gyrotropic Hall effect belongs, has recently enjoyed a revived interest. It is thus worthwhile to review the theoretical developments, after a brief reminder that in a realistic, impure system, the linear anomalous Hall effect \cite{NagaosaReview} stems from often comparable intrinsic (Berry curvature) and extrinsic (skew scattering and side jump) contributions. Disregarding the extrinsic effects, it was recently demonstrated \cite{SodemannFu2015} that the low-frequency nonlinear dc and ac current response in monochromatic field is related to the Berry curvature dipole and thus topologically quantized in three dimensions. An analogous quantization was proposed \cite{Morimoto2017CPVE} for the optical high-frequency regime of the dc response (the circular photogalvanic effect). In fact, all quantized nonlinear effects are diagrammatically associated with the same anomalous triangular diagram \cite{ParkerMoore2018}. Contrary to those previous works on nonlinear effects, we here do take extrinsic contributions into account and show their parametric importance. Still, we are able to demonstrate that the topologically quantized trace holds for the gyrotropic Hall effect in Weyl semimetals, even when extrinsic effects are taken into account. In passing, we remark that, upon reinterpretation of the ac field as a background drive and of the dc field constituting the probe, the gyrotropic Hall effect is furthermore related to the transport properties of topological Floquet materials~\cite{LindnerGalitski2011,KitagawaDemler2011,ChanRan2016,RudnerSong2018} for which, to the best of our knowledge, extrinsic effects have not been discussed.

The rest of the paper is organized as follows: in Section \ref{sec:phenomenology}, we summarize the phenomenology of current-induced Hall response and current-induced optical activity in gyrotropic metals. Section \ref{sec:results} contains the main results of this work. In Section \ref{sec:derivations}, we derive those results from the semiclassical Boltzmann equation. Section \ref{sec:applications} is devoted to the application to exemplary systems: Weyl semimetals and transition metal dichalcogenides under strain. We conclude in Section \ref{sec:discussion} with a final summary of the paper, and also provide numerous appendices with supplemental technical calculations.  


\section{Current-induced Hall response and optical activity in metals}\label{sec:phenomenology}

In this Section, we briefly recapitulate the phenomenology of the Hall response in metals, and summarize the magneto-optical effects that are related to it. Such phenomena, related to light polarization rotation in crystals stem from the presence of a Hall-like contribution in conductivity tensor of a material~\cite{LL8}:
\begin{equation}\label{eq:sigmaHall}
  \sigma^H_{ab}(\w)=\frac{e^2}{h}\epsilon_{abc} G_c(\w),
\end{equation}
where the pseudovector $\bm{G}$, in 3D, has dimensions of inverse length, and is odd under time reversal.

The typical origin of the polarization rotation phenomena lie in the presence of time-reversal breaking in a magnetically-ordered crystal. However, it has been noticed a while ago \cite{baranova1977,Ivchenko1978,Ivchenko1979} that in noncentrosymmetric crystals passage of a transport current can induce rotation of light polarization. This possibility is dictated by symmetry considerations: viewing $\bm G$ as a linear response to the transport current, one immediately concludes that it is odd with respect to the operation of time reversal, and the lack of an inversion center in the crystal allows $\bm G$ to be a pseudovector, even though the current density is a polar one. 

In this paper, we construct the theory of vector $\bm G$ in Eq.~\eqref{eq:sigmaHall} induced by passage of a transport current in metals with nontrivial band geometry, notably in Dirac and Weyl materials, in various frequency ranges. Such response can be quantified by writing the antisymmetric (Hall) part of the conductivity tensor as

\begin{align}\label{eq:sigmaHall_results}
\sigma^H_{ab}(\w)=\Lambda_{abc}(\w)E^{0}_{c},
\end{align}
where $\EE^{0}$ is the transport dc electric field that drives the transport current through the sample. Evidently, third-rank tensor $\Lambda_{abc}$ is antisymmetric with respect to the first pair of indices, hence is dual to a second-rank pseudotensor $\lambda_{ab}, \Lambda_{abc}=\epsilon_{abd}\lambda_{dc}$. We will refer to $\lambda_{ab}$ as the gyrotropic Hall tensor. Using this tensor, we can relate vector $\bm G$ of Eq.~\eqref{eq:sigmaHall} to the transport electric field: 
\begin{align}
  G_a(\w)=\frac{h}{e^2}\lambda_{ab}(\w)E^{0}_{b}.
\end{align}
It is clear that the knowledge of either tensor $\Lambda_{abc}$, or the gyrotropic Hall tensor $\lambda_{ab}$ is sufficient to fully describe the current-induced Hall response of a chiral metal, and the GHE. 

Before presenting the semiclassical kinetic theory, we would like to outline the phenomenology of optical polarization rotation in metals for future reference. For simplicity and clarity of presentation, we assume that the Hall-like response defined by Eq.~\eqref{eq:sigmaHall} is the only nondiagonal part of the conductivity tensor. Various aspects of the optics of crystals can be added to consideration along the textbook lines \cite{LL8}. The refractive indices for the two circular polarizations of light ($R,L$) are given by \cite{Argyres1955,Bader2000}
\begin{equation}\label{eq:refractive}
  n_{L,R}=n_\w\left(1\mp\frac12 \hat{\bm q}\cdot {\bm g}\right),\,\,\,{\bm g}=\frac{1}{\e_0\e_\w\w}\frac{e^2}{h}{\bm G}.
\end{equation}
Here $\hat{\bm q}$ is the direction of the wave vector, and $\e_\w=n_\w^2$ are the background dielectric function and the refractive index of the material. 

In this paper, we are interested in magneto-optical phenomena induced by current passage through a metal with nontrivial band geometry, focusing on Kerr and Faraday polarization rotation angles. Since both of these are determined by the (difference in) refractive indices for the two circular polarizations of light, Eq.~\eqref{eq:refractive} provides all necessary information to consider magneto-optical phenomena. 

For propagation along the $\bm G\equiv (0,0,G)$ vector, which yields the largest rotation according to Eq.~\eqref{eq:refractive}, the Kerr rotation angle is given by
\begin{align}\label{eq:Kerr}
\theta_{K}=\Im \frac{n_R-n_L}{n_Rn_L-1}\simeq \Im \frac{\sigma_{xy}}{\e_0\w \sqrt{\e_\w}(\e_\w-1)}.
\end{align}
Here $\sigma_{xy}=e^2 G/h $ is the Hall conductivity in the plane perpendicular to vector $\bm G$, and the second approximate equality holds for small $\sigma_{xy}$.

In turn, the Faraday rotation per unit thickness of material is 
\begin{align}\label{eq:Faraday}
\rho_F=\frac{\w}{2c}\Re (n_L-n_R)\simeq-\frac{1}{2\e_0c}\Re \frac{\sigma_{xy}}{ \sqrt{\e_\w}}.
\end{align}
Note the polarization rotation angles are defined as positive for rotation from the $x$-axis to $y$-axis, for light propagating along $-z$-axis. It is also noteworthy that in the absence of absorption, $\Im (\e_\w)=0$, the Kerr and Faraday rotations are determined by the imaginary and real parts of the Hall conductivity, respectively. 

In this paper, besides three-dimensional materials, we will consider the GHE in 2D systems. Therefore, we present the corresponding expressions for the Kerr ($\theta^{\textrm{2D}}_{K}$) and Faraday ($\theta^{\textrm{2D}}_{F}$) rotation angles for a 2D layer assumed to be in vacuum: 
\begin{align}\label{eq:2Dangles}
    &\theta^{\textrm{2D}}_K\simeq -\Re\frac{\sigma^{\textrm{2D}}_{xy}}{\sigma^{\textrm{2D}}_{xx}},\quad 
    &\theta^{\textrm{2D}}_F\simeq \Re \frac{\sigma^{\textrm{2D}}_{xy}}{2\epsilon_0c}.
\end{align}
It is assumed above that the 2D conductivities satisfy $\sigma^{\textrm{2D}}_{xy}\ll \sigma^{\textrm{2D}}_{xx}$, and $\sigma^{\textrm{2D}}_{xx}/\epsilon_0c\ll1$. Eqs.~\eqref{eq:2Dangles} show that the Faraday rotation angle is a direct measure of the Hall response in a 2D layer. However, the value of the Kerr rotation angle is larger by the factor of $\epsilon_0c/\sigma^{\textrm{2D}}_{xx}\gg1$, hence Kerr rotation is a much more sensitive probe (in particular, of the GHE). 

We would like to finish this Section with a brief comment on the relation between the GHE and the phenomenon of current-induced magnetization, also known as the Rashba-Edelstein effect, or the kinetic magneto-electric effect~\cite{Levitov1985,Edelstein1990,Rou2017,Yoda2018}. It is tempting to interpret the GHE as a type of anomalous Hall effect that stems from the current-induced magnetization, rather than one present in the ground state of a magnetic sample. Indeed, one can eliminate the electric field from the linear relation~\eqref{eq:sigmaHall_results}, and a similar relation for the current-induced magnetization, thereby relating the hall conductivity tensor directly to the magnetization. However, such a relation  between the GHE and current-induced magnetization is not particularly deep, since it does not reveal the physical origin of the GHE. Indeed, as it will become apparent below, the GHE and kinetic magneto-electric effect stem from different geometric objects. While the magnetization  is naturally related to the existence of spin and orbital magnetic moments of electrons, the GHE comes from the anomalous velocity of charge carriers. Therefore, the GHE and the kinetic magneto-electric effect should be considered as two separate phenomena, both of which are caused by band geometry, and hence appear simultaneously in the presence of a transport current.


\section{Summary of results}\label{sec:results}

As explained in Section~\ref{sec:phenomenology}, the GHE is described with a single second-rank gyrotropic Hall (pseudo-) tensor, $\lambda_{ab}$, or, equivalently, with the nonlinear ac-dc Hall tensor $\Lambda_{abc}$. In this section, we summarize the result obtained in subsequent sections for these two quantities. 

\subsection{Low-frequency result for gyrotropic Hall tensor}

\begin{figure}
    \centering
    \includegraphics[scale=.28]{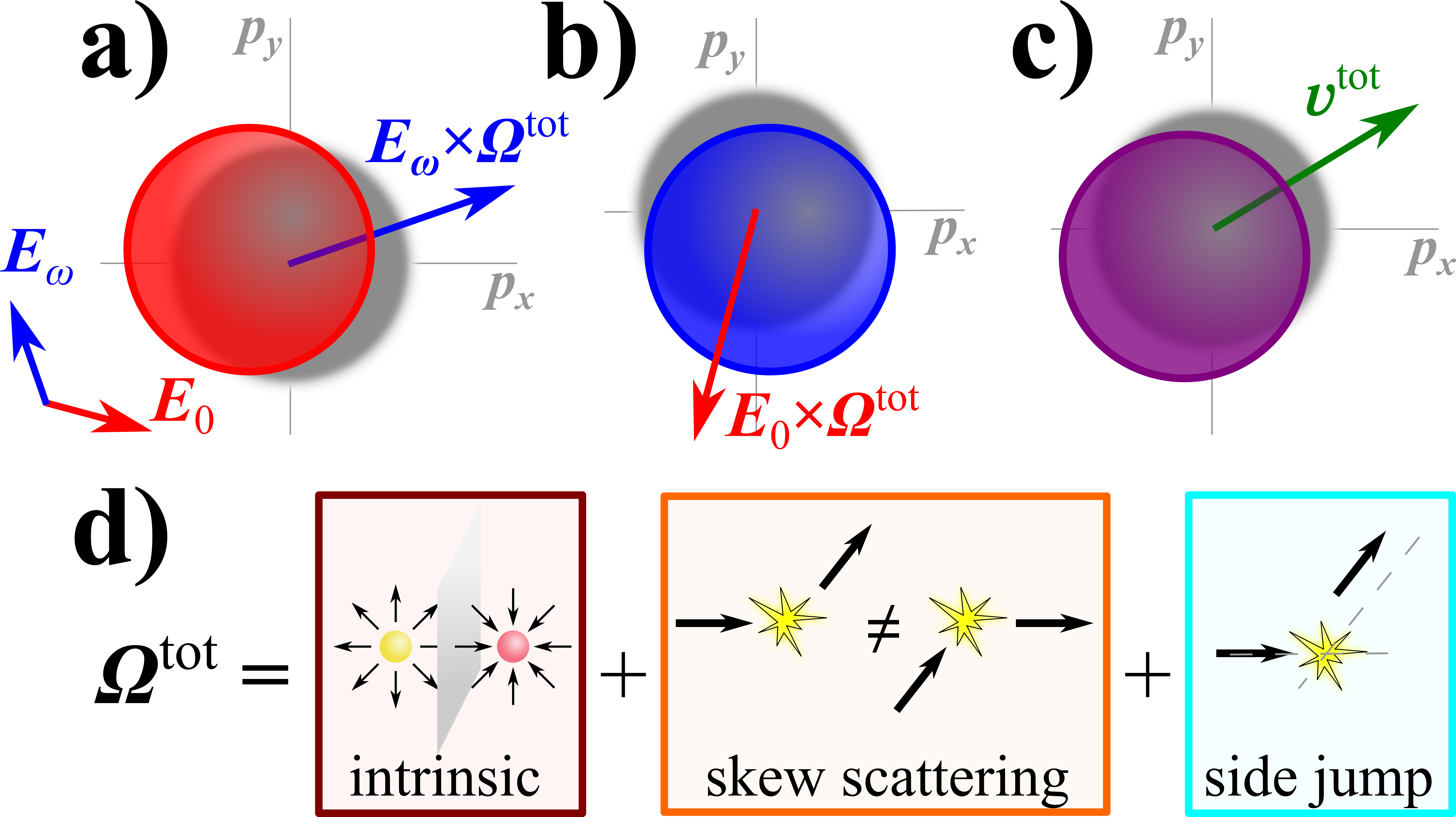}
\caption{Cartoon of contributions to the gyrotropic Hall tensor, per Eqs. ~\eqref{eq:LambdaFullResult} and \eqref{eq:lambdatensor}. Panels (a) and (b) represent the evaluation of the anomalous velocity $\EE \times \boldsymbol \Omega$ (with either $\EE^0$ or $\EE^\omega$) in the presence of a Fermi surface shifted by $\tau e \EE^0$ ($\tau_\omega e \EE^\omega$). In contrast, in panel (c) the zero field velocity is evaluated employing the nonlinear shift of momenta by $\tau e \EE^0 + \tau_\omega e \EE^\omega$. Lastly in panel (d) we illustrate extrinsic processes due to skew scattering and side jump anomalous events that may have contributions of the same order as the intrinsic effect.}
    \label{fig:Cartoon}
\end{figure}

We explicitly show that in the presence of disorder, the gyrotropic Hall tensor has both intrinsic and extrinsic contributions. The intrinsic part is determined by the Berry curvature dipole moment, while the extrinsic ones are governed by the skew scattering and side jump processes, familiar from the theory of anomalous Hall effect \cite{SinitsynReview,NagaosaReview}. The physical meaning of those terms becomes most apparent in the definition $j_a = \Lambda_{abc} E^{\omega}_{b} E^{0}_{c}$, where we  express $\Lambda_{abc}$ as a Fermi surface integral
\begin{eqnarray}
\Lambda_{abc} &=& e^3 \int_{\pp} \tau \epsilon_{abd} \Omega_d^{\rm tot} v_c \partial_{\ve_\pp}f_{\mathrm{eq}}(\ve_{\pp}) \notag \\
&+&e^3\int_{\pp}  \tau_\omega \epsilon_{acd} \Omega_d^{\rm tot} v_b \partial_{\ve_\pp}f_{\mathrm{eq}} (\ve_{\pp})\notag \\
&-&e^3 \int_{\pp} \tau \tau_\omega v^{\rm tot}_a \partial_c v_b \partial_{\ve_\pp}f_{\mathrm{eq}} (\ve_{\pp}). \label{eq:LambdaFullResult}
\end{eqnarray}
Here, antisymmetrization in $a \leftrightarrow b$ is implied, $\tau$ is the elastic scattering time, and we defined 
\begin{equation} 
\tau_\omega = \frac{\tau}{1- i \omega \tau}. 
\end{equation}
The  total Berry curvature $\boldsymbol{\Omega}^{\rm tot}$ and total velocity $\vv^{\rm tot}$ will be introduced shortly in Sec.~\ref{sec:Summary:Defs}.
The first line represents the full ac anomalous velocity $\vv^{\rm anom} = e (\EE^\omega \times  \boldsymbol{\Omega}^{\rm tot})$ evaluated with the nonequilibrium distribution function due to the momentum shift $\ve_{\pp} \rightarrow \ve_{\pp - \tau e \EE^0}$, see Fig.~\ref{fig:Cartoon}(a). The second line is the same, but the role of ac and dc fields is reversed, Fig.~\ref{fig:Cartoon} (b). Finally, the last line is the evaluation of the total velocity taking into account the nonlinearity of the dispersion relation, see Fig. \ref{fig:Cartoon}(c). 

\subsection{Total velocity and Berry curvature}
\label{sec:Summary:Defs}

The total velocity entering Eq.~\eqref{eq:LambdaFullResult} 
\begin{equation}
    \vv^{\rm tot} = \boldsymbol{\nabla}_{\pp} \ve_{\pp} + \frac{\tau_\omega}{\tau}(\tau_\omega \bm{A}^{\rm sk}+2 \vv^{\rm sj}) \label{eq:vtot}
\end{equation}
contains corrections to the group velocity originating from skew-scattering (Sec.~\ref{sec:skew}) and side jump (Sec.~\ref{sec:sidejump}) mechanisms.  

 The expressions for the skew acceleration $\bm A^{\textrm{sk}}_\pp$ and side jump accumulation $\vv^{\textrm{sj}}_\pp$ involve the symmetric and antisymmetric parts of the impurity scattering probability $w^{(S,A)}_{\pp\pp'}$ from $\pp'$ to $\pp$, see Eqs.~  \eqref{eq:skewcollisionintegral} and \eqref{eq:symantisym}, as well as the coordinate shift of the electron in the same transition, $\d\rr_{\pp\pp'}$, Eq.~\eqref{eq:sidejump}. The physical meaning of $\bm A^{\textrm{sk}}_\pp$ and $\vv^{\textrm{sj}}_\pp$ becomes clear from their respective microscopic definitions given in Eqs. \eqref{eq:skewacceleration} and \eqref{eq:sjvelocity} respectively: $\vv^{\textrm{sj}}_\pp$ is a disorder-induced velocity of a wave packet due to the accumulation of the side jump events; in turn, $\bm A^{\textrm{sk}}_\pp$ is a disorder-induced acceleration of a wave packet due to the velocity change accumulation upon skew scattering.

The total Berry curvature is defined as
\begin{equation}\label{eq:BerryTotal}
    \boldsymbol \Omega^{\rm tot} = \boldsymbol{\nabla}_{\pp} \times \left[i \braket{u_{\pp} \vert \boldsymbol \nabla_{\pp} u_\pp} + \frac{\tau}{2} \vv^{\rm tot}\right].
\end{equation}
The intrinsic part of the curvature describes the geometry of Bloch wavefunctions $\vert u_{\pp} \ket$ over the Brillouin zone torus. It follows from the expression of the linear Hall response
\begin{equation}
    \sigma_{ab} = - e^2 \epsilon_{abc} \int_{\pp} \Omega_c^{\rm tot} f_{\mathrm{eq}}(\ve_\pp),
\end{equation}
that the geometric meaning of $\boldsymbol \Omega^{\rm tot}$ is to characterizes the fiber bundle of wave functions of the entire impure system for twisted boundary conditions \cite{NiuWu1985}.
We remark that we added the contributions from side jump accumulation and anomalous distribution \cite{NagaosaReview}, hence the factor of 2 in front of $\vv^{\rm sj}$ in Eq.~\eqref{eq:vtot}.

\subsection{Topological quantization and frequency dependence}

From Eq.~\eqref{eq:LambdaFullResult}, it follows that the gyrotropic Hall tensor is defined by the dipole moment of the total Berry curvature in the Brillouin zone, $D_{ab}^{\rm tot}$, which we generalize as compared to the theory of the intrinsic photogalvanic effect~\cite{Deyo2009,SodemannFu2015}:
\begin{align}\label{eq:Dtensor}
  D_{ab}^{\rm tot}=\int_{\pp} \Omega_b^{\rm tot}\p_{a} f_{\mathrm{eq}}.
\end{align}
Under the assumption of momentum independent scattering times, $\lambda_{ab}$ can be expressed in terms of $D_{ab}^{\rm{tot}}$, 
\begin{align}\label{eq:lambdatensor}
  \lambda_{ab} &=e^3\tau \left[\left(1+\frac{\tau_\w}{2\tau}\right)D^{\rm tot}_{ba}-\frac{\tau_\w}{2\tau} \d_{ab} \Tr D^{\rm tot}\right] \notag \\
  &- e^3 \frac{\tau_\omega \tau}{2} \int_{\pp} [(\vv^{\rm tot} \times \boldsymbol{\nabla}_\pp)_a v_b] \partial_{\e_\pp}f_{\mathrm{eq}}  (\ve_\pp) .
\end{align}
It is noteworthy that in 3D materials without Weyl points, or in 2D materials, the tensor $D^{\rm tot}$ is traceless, and $\lambda$ inherits that property. This statement stems from an observation that the integral 
\begin{align}\label{eq:Berrycharge}
\frac1{2\pi}\int_{\pp} (\bm \Omega\cdot\vv)\, \p_{\ve_\pp} f_{\mathrm{eq}}=-\frac1{2\pi}\int_{{\textrm{fs}}} d\bm S_{\textrm{fs}}\cdot \bm \Omega_{\textrm{fs}},
\end{align}
where $d\bm S_{\textrm{fs}}$ is the directed Fermi surface element, measures the total Berry monopole charge inside a Fermi pocket. This quantity obviously vanishes in the absence of Berry monopoles, \textit{i.e.} in the absence of Weyl points. When such points are present in 3D metals, it is helpful to reinstate an explicit summation over the Fermi surface sheets in the expression for $\Tr \lambda$: 
\begin{align}\label{eq:Trlambda}
  \Tr \lambda=\sum_{\textrm{fs}}\frac{2\pi i e^3\w \tau^2_{\textrm{fs}}}{(1-i\w\tau_\textrm{fs})}Q_\textrm{fs},
\end{align}
in which $Q_\textrm{fs}$ is the Berry monopole charge contained inside a Fermi surface (note that the Berry charge is the same for both electronic and hole Fermi surfaces around a given Weyl point, since both Berry curvature and the direction of the ``outer'' normal are opposite for them). Therefore, we conclude that in the relaxation time approximation, the trace of tensor $\lambda_{dc}$ that defines the gyrotropic Hall effect is of topological origin and it can be readily seen that extrinsic contributions drop in Eq.~\eqref{eq:Trlambda}.

Compared to the usual case of the anomalous Hall effect \cite{NagaosaReview}, the most peculiar feature of the gyrotropic Hall effect is its frequency dependence, which allows to separately discuss and experimentally distinguish intrinsic from extrinsic mechanisms of GHE. Indeed, in the high-frequency limit, $\w\tau\gg 1$, or equivalently $|\tau_\w|/\tau\ll 1$, skew scattering contributions are suppressed by $1/(\omega \tau)^2$. In this limit
\begin{equation}\label{eq:LargeFreq}
    \lambda_{ab} \simeq {e^3}{\tau} D_{ba}^{\rm int} + \frac{i e^3}{\omega}\left [ \frac{D_{ba}^{\rm int} -\d_{ab}\tr D^{\rm int}}{2} + D_{ba}^{\rm sj} \right],
\end{equation}
where the superscript ``int'' (``sj'') indicates that only the intrinsic (side jump) contribution to the Berry curvature, Eq.~\eqref{eq:BerryTotal}, is kept. 

In the opposite limit of low frequencies, $\w\tau\ll 1$, the distinction between the intrinsic and extrinsic mechanisms is as hard as in the usual AHE case. Note that the extrinsic correction to the Berry curvature, Eq.~\eqref{eq:BerryTotal}, can be of comparable size to the intrinsic one, and can even reverse the sign of their totality, see Section~\ref{sec:TMD} for an example. Even more drastically, as in the case of the AHE, in the limit of rare strong impurities, the skew scattering contribution dominates the effect. We expect both statements to also hold in the case of the dc nonlinear Hall effect \cite{Deyo2009,SodemannFu2015,ma2018}.

\subsection{Large frequencies}
All preceding considerations have been restricted to the low-frequency regime, in which the ac frequency is small as compared to relevant band splittings. In the opposite case of optical frequencies, provided that the crystal is sufficiently clean and the bands are well resolved, mechanisms related to disorder scattering are not efficient, see Eq.~\eqref{eq:LargeFreq}. For a short discussion of the intrinsic high frequency contributions we refer to Sec.~\ref{sec:HighFreq}, and summarize here the result for a generic two-band model. An interpolating function which smoothly connects the low-frequency result with the case of resonant optical frequencies is obtained by introducing
\begin{equation}\label{eq:BerryOmega}
    \boldsymbol \Omega^{\rm tot}_{\omega} = i \frac{\ve_{n \bar n}^2}{\ve_{n \bar n}^2 + \tau_\omega^{-2}} \boldsymbol{\nabla}_{\pp} \times \braket{u_{\pp} \vert \boldsymbol \nabla_{\pp} u_\pp} + \frac{\tau}{2} \boldsymbol{\nabla}_{\pp} \times \vv^{\rm tot}
\end{equation}
and redefining $D_{ab}^{\rm tot}$ by replacing $\boldsymbol \Omega^{\rm tot} \rightarrow \boldsymbol \Omega^{\rm tot}_{\omega}$ in Eq.~\eqref{eq:Dtensor}. Here, $\ve_n \equiv \ve_n (\pp)$ is the dispersion in band $n$ to which the Fermi surface under consideration belongs, and $\ve_{n\bar n} = \ve_n - \ve_{\bar n}$ measures the energy splitting to the second band $\bar n$ at the same momentum. Note that at large frequencies $\tau_\omega^{-1} \simeq - i \omega$ implying a resonance at $\vert \ve_{n \bar n} \vert = \w$.


\section{Kinetic theory of gyrotropic Hall effect in metals}\label{sec:derivations}
In this Section, we consider the phenomenon of the current-induced optical activity (GHE) in detail. To this end, we calculate the effective Hall-like response, Eq.\eqref{eq:sigmaHall}, induced by a dc (transport) current, in a generic noncentrosymmetric crystalline metal. It has momentum-space Bloch Hamiltonian $\hat h_\pp$, with an eigensystem that contains Bloch eigenstates $|u_{n\pp}\ket$, and energies $\ve_{n\pp}$:
\begin{align}
\hat h_\pp |u_{n\pp}\ket=\ve_{n\pp}|u_{n\pp}\ket.
\end{align}
In what follows we will suppress the band index $n$, with the understanding that it can be reinstated simply by adding summation over it to momentum-space integrals over $\pp$.

Given the materials of interest -- Dirac and Weyl semimetals -- we will restrict the ranges of optical excitation frequencies to the two most interesting one: the low-frequency regime, in which the optical frequency is small compared to relevant band splittings, $\w\ll\ve_g$; the resonant regime, in which the optical frequency is large compared to any intraband frequency (\textit{e.g.} the inverse transport time), and is close to an interband resonance.

In principle, both regimes of low and high frequencies can be considered based on the multiband quantum kinetic equation. However, here we choose to use the formalism of Boltzmann kinetic equation with semiclassical corrections taken into account for the low-frequency regime due to its physical transparency. The high-frequency regime must be considered within the multiband quantum kinetic equation. 

In the limit of small excitation frequencies, $\omega,1/\tau\ll \ve_g$, the finite-frequency response in the presence of a dc current can be obtained from the conventional Boltzmann kinetic equation, with semiclassical corrections taken into account \cite{XiaoNiu}. Such treatment neglects $O(\w/\ve_g)$ contributions. Further, Weyl semimetals with not too high doping levels are necessarily semiconductors with many valleys, hence have several Fermi pockets. In what follows, we neglect the inter-valley scattering, since the corresponding rate is typically small compared to the intra-valley one, and in the present problem does not bring any new physics.

The technical task we are facing below is to solve  coupled integro-differential kinetic equations to a nonlinear order in applied field. This is an extremely challenging task even with above specified simplificatins, so a few additional comments are in order in relation to kinetic scheme outlined here. (i) We will work only up to order $O(E^0\cdot E^\w)$ in applied dc, $\EE^0$, and ac, $\EE^\w$, electric fields. (ii) We will work in the simplified model case with the single relaxation time approximation (RTA). In other words, we disregard the fact that angular harmonics of the distribution function may decay on different time scales, although we expect these time scales to be parametrically equivalent. (iii) We further assume that relaxation time is momentum independent, namely constant RTA model. While such approximation is inadequate to capture thermoelectric and thermomagnetic responses, it will be sufficient to describe GHE which does not require particle-hole asymmetry. We stress that the physics of GHE does change if relaxation time is energy dependent, but the change is qualitative and does not bring any new conceptual features to the form of GHE tensor. (iv) We treat collision terms perturbatively by iterations thus assuming weak and smooth impurity potential. (v) We ignore interference effects between different scattering channels contributing to gyrotropic Hall response as these are higher order corrections.     

\subsection{Intrinsic mechanism}

In the present case, the electric field has bi-harmonic time-dependence, $\EE(t)=\EE^0+\EE^\w e^{-i\omega t}$. We neglect its spatial dependence. The kinetic equations for the dc, $f^{0}$, and ac, $f^{\w}$, nonequilibrium components of the distribution function in a given valley (valley index suppressed), are 
\begin{align}\label{eq:kineqs}
  e\EE^0\p_\pp f_{\mathrm{eq}}=I\{f_\pp^0\},\nonumber\\ 
  -i\w f^{\w}_\pp+e\EE^\w\p_\pp (f_{\mathrm{eq}}+f^0_\pp)+e\EE^0\p_\pp f^\w_\pp=I\{f_\pp^\omega\},
\end{align}
where the integral operator $I\{f_\pp\}$ encapsulates collision terms. We assume that the main source of scattering is disorder for which the usual collision integral reads 
\begin{equation}\label{eq:skewcollisionintegral}
  I\{f_\pp\}=-\int_{\pp'} (w_{\pp'\pp}f_{\pp}-w_{\pp\pp'}f_{\pp'})\d(\ve_{\pp}-\ve_{\pp'}),
\end{equation}
with a scattering rate that generically contains parts that фку both symmetric and antisymmetric with respect to the interchange of the initial and final states
\begin{align}
w_{\pp\pp'}=w^S_{\pp\pp'}+w^A_{\pp\pp'}, \quad 
w^{S,A}_{\pp\pp'}=\pm w^{S,A}_{\pp'\pp}. \label{eq:symantisym}
\end{align}
Kinetic equations~\eqref{eq:kineqs} must be supplemented with an expression for the current
\begin{equation}\label{eq:intrinsiccurrent}
\jj=e\int_\pp \left(\bm{\nabla}_\pp\varepsilon_\pp+e\bm{\Omega}_\pp\times\bm{E}\right)f_\pp   
\end{equation}
that contains both band and anomalous velocity terms. Note that both the energy-conserving $\d$-function in the impurity collision integral, Eq.~\eqref{eq:skewcollisionintegral}, as well as the expression for the current, Eq.~\eqref{eq:intrinsiccurrent}, must be modified to take into account side jump processes. This will be described in more detail in Sec.~\ref{sec:sidejump}.

For the purpose of calculating the intrinsic contribution to the antisymmetric part of the conductivity tensor, only the anomalous velocity current due to the Berry curvature of the band structure, $\bm \Omega_\pp$ (valley index suppressed) needs to be taken into account. Such anomalous velocity currents are linear in electric fields, hence it suffices to solve kinetic equations~\eqref{eq:kineqs} to linear order in transport and optical fields. We denote such solutions for the distribution function as $f^{\EE^0}$ and $f^{\EE^\w}$, which describe response to the transport and optical fields, respectively. We can write the corrections to the equilibrium distribution function in the standard form: 
\begin{align}\label{eq:linearizeddf}
f^{\EE^0}_\pp=-\tau e\EE^0\p_\pp f_{\mathrm{eq}},\quad 
f^{\EE^\w}_\pp=-\tau_\w e\EE^\w\p_\pp f_{\mathrm{eq}},
\end{align}
where the elastic scattering time is defined by symmetric part of the scattering probability in a usual fashion
\begin{equation}
\tau^{-1}=\int_{\pp'}w^S_{\pp\pp'}(1-\cos\theta_{\pp\pp'})\delta(\e_\pp-\e_{\pp'}).
\end{equation}
The anomalous current oscillating at the frequency of the optical field is then given by
\begin{align}\label{eq:anomalouscurrent}
  \jj^{\EE^0\EE^\w}=e^2\int_{\pp} (\bm \Omega_\pp\times \EE^\w) f^{\EE^0}_\pp+e^2\int_{\pp} (\bm \Omega_\pp\times\EE^0) f^{\EE^\w}_\pp.
\end{align}
The total anomalous current is given by summing the current in Eq.~\eqref{eq:anomalouscurrent} over all valleys. Substituting the solutions of Eqs.~\eqref{eq:kineqs} into Eq.~\eqref{eq:anomalouscurrent} for the current, and separating the antisymmetric part of the corresponding optical conductivity tensor, we obtain the intrinsic contribution to the GHE, $\sigma_{ab}^\w=\Lambda^{\textrm{int}}_{abc} E^{0}_{c}$ (summation over repeated indices is implied), with 
  \begin{align}
  \Lambda^{\textrm{int}}_{abc}=e^3\! \int_{\pp} \tau \left[\epsilon_{abd}v_c+\frac{\tau_\w}{2\tau}(\epsilon_{acd}v_b - \epsilon_{bcd}v_a) \right]\Omega_d\p_{\ve_\pp}f_{\mathrm{eq}}, \label{eq:Lambdaabc}
\end{align}
where we introduced the band velocity $\vv_\pp\equiv\p_\pp\ve_\pp$.
For $\Lambda_{abc}$ given by Eq.~\eqref{eq:Lambdaabc}, the gyrotropic Hall (pseudo)tensor dual to it is given by
\begin{align}\label{eq:lambda}
  \lambda^{\textrm{int}}_{dc}=e^3\int_{\pp}\tau \left[\left(1+\frac{\tau_\w}{2\tau}\right)\Omega_dv_c-\frac{\tau_\w}{2\tau}\left(\bm \Omega\cdot\bm v\right)\d_{dc}\right]\p_{\ve_\pp}f_{\mathrm{eq}}.
\end{align}
Note that in the constant relaxation time approximation (for a given Fermi surface), the gyrotropic Hall tensor is defined by the dipole moment of the Berry curvature in the Brillouin zone, $D_{ab}$ familiar from the theory of photogalvanic effect~\cite{Deyo2009,SodemannFu2015}:
\begin{align}\label{eq:DtensorINT}
  D_{ab}=\int_{\pp} \Omega_b\p_{a} f_{\mathrm{eq}}.
\end{align}

In terms of $D_{ab}$, and assuming momentum independent scattering rates $\lambda_{ab}$ is given by 
\begin{align}\label{eq:lambdatensorINT}
  \lambda^{\textrm{int}}_{ab}=e^3\tau \left[\left(1+\frac{\tau_\w}{2\tau}\right)D_{ba}-\frac{\tau_\w}{2\tau}\Tr D \d_{ab}\right].
\end{align}
In particular, the trace of $\lambda_{ab}$ is given by 
\begin{align}
  \Tr \lambda^{\textrm{int}}=-i e^3\w \tau\tau_\w\Tr D.
\end{align}

\subsection{Skew scattering mechanism}
\label{sec:skew}

In addition to the intrinsic contribution, the first extrinsic part of the anomalous transport is provided by the skew scattering mechanism. In the broader sense of AHE \cite{SinitsynSinova2007}, this contribution accounts for the asymmetry of the scattering probability $w_{\pp \leftarrow \pp'} \neq w_{\pp' \leftarrow \pp}$ which may result from strong impurities treated beyond the Born approximation \cite{smit1955}, spin-orbit-active impurities \cite{Milletari2016}, as well as from virtual \cite{SinitsynSinova2007} and diffractive \cite{AdoTitov2015,KonigLevchenko2016} scattering off two-impurity complexes. The skew scattering is defined by the antisymmetric part of the scattering rate. The probability conservation dictates the following property of $w^A_{\pp\pp'}$\cite{Belinicher1980}:
\begin{align}\label{eq:unitarity}
\int_{\pp'} w^A_{\pp\pp'}\d(\ve_\pp-\ve_{\pp'})=0.
\end{align}

In the presence of the skew scattering, the kinetic equations~\eqref{eq:kineqs} are written as 
\begin{align}\label{eq:skewkineqs}
    e\EE^0\p_\pp f_{\mathrm{eq}}=-\frac{1}{\tau}(f^{0}_\pp-\langle f^{0}_\pp\rangle)+\int_{\pp'}W^A_{\pp\pp'}f^{\EE^0}_{\pp'},\nonumber\\
  -i\w f^{\w}_\pp+e\EE^\w\p_\pp (f_{\mathrm{eq}}+f^0_\pp)+e\EE^0\p_\pp f^\w_\pp=\nonumber \\ 
  -\frac{1}{\tau}(f^{\w}_\pp-\langle f^{\w}_\pp\rangle)+\int_{ \pp'}W^A_{\pp\pp'}f^{\EE^\w}_{\pp'},
\end{align}
where $W^A_{\pp\pp'}\equiv w^A_{\pp\pp'}\d(\ve_\pp-\ve_{\pp'})$.
Evidently, the skew scattering affects both linear and nonlinear responses of a metal to electric field. Since the solution to the linear problem is used as an input for the nonlinear one, we first describe the former. 

The corrections to the distribution function linear both in the electric fields, and in the skew scattering rate are obtained by inserting the solutions without skew scattering, Eqs.~\eqref{eq:linearizeddf}, into the integrals on the right hand side of Eqs. \eqref{eq:skewkineqs}, whereby they become additional generation terms linear in the skew scattering rate. Balancing these generation terms with the isotropic scattering rate, or the time derivative, one obtains an additional change in the distribution function: 
\begin{align}\label{eq:skewdflinear}
&\d f^{\EE^0}_{\pp}=-\tau^2e\EE^{0}\int_{ \pp'}W^A_{\pp\pp'}\p_{\pp'} f_{\mathrm{eq}},\nonumber\\
 &\d f^{\EE^\w}_\pp=-\tau_\w^2e\EE^{\w}\int_{\pp'}
 W^A_{\pp\pp'}\p_{\pp'} f_{\mathrm{\mathrm{eq}}}.
\end{align}

Turning to the nonlinear case, we first present the solution in the case without the skew scattering. The nonlinear ac correction to the distribution function is found from 
\begin{align}\label{eq:nonlinkineq}
  -i\w f^{\EE^0\EE^\w}_{\pp}+e\EE^\w\p_\pp f^{\EE^0}_{\pp}+e\EE^0\p_\pp f^{\EE^\w}_{\pp}=\nonumber \\ -\frac{1}{\tau}(f^{\EE^0\EE^\w}_{\pp}-\langle f^{\EE^0\EE^\w}_{\pp}\rangle).
\end{align}
Even though the generation terms for $f^{\EE^0\EE^\w}_{\bm{p}}$ - the second and third terms in the left hand side of Eq.~\eqref{eq:nonlinkineq} -  may have nonzero angular averages, we do not include an additional inelastic collision integral to relax them, since the ac nature of $f^{\EE^0\EE^\w}_{\bm{p}}$ allows stabilization of the generation terms by the time derivate (the first term in the left hand side of Eq.~\eqref{eq:nonlinkineq}). Further, since it can be shown that $\langle f^{\EE^0\EE^\w}_{\pp}\rangle$ does not make a contribution to the Hall-like response, we can write   
\begin{align}\label{eq:nonlineardistribution}
f^{\EE^0\EE^\w}_{\pp}=e^2 E^{0}_{c}E^{\w}_{b}
\left(\tau+\tau_\w\right)\tau_\w \p_b\p_c f_{\mathrm{eq}}.
\end{align}

The skew scattering contributes to the nonlinear response in two ways: (i) the linear response corrections to the distribution function, Eqs.~\eqref{eq:skewdflinear}, are substituted into the electric-field drive terms in the second of Eqs.~\eqref{eq:skewkineqs}, whereby turning them into additional nonlinear generation terms; (ii) the nonlinear correction to the distribution function, Eq.~\eqref{eq:nonlineardistribution}, is substituted into the antisymmetric part of the collision integral in the second of Eqs.~\eqref{eq:skewkineqs}, yielding another generation term linear in the skew scattering rate. 

With the same provisions as for Eqs.~\eqref{eq:skewdflinear}, and integrating by parts as needed, the nonlinear correction to the distribution function, which is linear in the skew scattering rate can be written as 
\begin{align}\label{eq:nonlinskew}
  \delta f^{\EE^0\EE^\w}_{\pp}=e^2 E^{0}_{c} E^{\w}_{b} \int_{ \pp'}\left[\tau^2\tau_\w\p_b W^A_{\pp\pp'}\p'_c f_{\mathrm{eq}}\right. \nonumber \\ 
  \left.-\tau\tau_\w^2\p'_b W^A_{\pp\pp'} \p'_c f_{\mathrm{eq}}+
 \tau_\w^3 (\p_c-\p'_c) W^A_{\pp\pp'}\p'_{b} f_{\mathrm{eq}}\right].
\end{align}
Again as we employ constant RTA then upon integration by parts derivatives do not act on $\tau$. The first term in square brackets on the right hand side of this expression gives a symmetric contribution to the conductivity tensor. This can be seen by noting that $\vv_\pp=\p_\pp \ve_\pp$, and doing an integration by parts in the $\pp$-integral \cite{note_symmetric}.

To write the contribution to the conductivity tensor stemming from Eq.~\eqref{eq:nonlinskew}, it is natural to associate the ``skew acceleration'' with the skew scattering process \cite{Rou2017}, 
\begin{equation}\label{eq:skewacceleration}
  \bm A^{\textrm{sk}}_\pp=\int_{\pp'}W^{\textrm{A}}_{\pp\pp'}
(\p_{\pp}\e_{\pp}-\p_{\pp'}\e_{\pp'}),
\end{equation}
which describes the rate of change of the carrier velocity due to skew scattering collisions. Note that the integral containing $\p_\pp \ve_\pp$ vanishes due to property~\eqref{eq:unitarity} of $W^{\textrm{A}}_{\pp\pp'}$. We kept it in Eq.~\eqref{eq:skewacceleration} for clarity and to emphasize the physical meaning of $\bm A^{\textrm{sk}}_\pp$.

Calculating the current that corresponds to the distribution function~\eqref{eq:nonlinskew}, and taking the antisymmetric part of the corresponding conductivity tensor, one obtains
\begin{align}\label{eq:Lambdaskew}
 \Lambda^{\textrm{sk}}_{abc}=\frac12e^3\tau\tau_\w^2\int_ {\pp}\left[\left(1+\frac{\tau_\w}{\tau}\right)(\p_a A^{\textrm{sk}}_b-\p_b A^{\textrm{sk}}_a)v_c \right.\nonumber \\ \left.-
 \frac{\tau_\w}{\tau} (A^{\textrm{sk}}_a\p_c v_b-A^{\textrm{sk}}_b\p_c v_a)\right]\p_{\ve_\pp}f_{\mathrm{eq}}.
\end{align}
The same expression can be rewritten in a different form that makes natural connection to the structure of the intrinsic contribution. Indeed, within constant RTA the above expression is identical to  
\begin{align}
\Lambda_{abc}^{\rm sk} = e^3 \tau \int_{\pp} \left \lbrace \left [ \epsilon_{abd} v_c + \frac{\tau_\omega}{2 \tau} (\epsilon_{acd} v_b  - \epsilon_{bcd} v_a) \right ]\Omega_d^{\rm sk} \right. \nonumber \\ \left. + \frac{\tau_\omega^3}{2 \tau} (\partial_c v_a A^{\textrm{sk}}_b  - \partial_c v_b A^{\textrm{sk}}_a) \right \rbrace \p_{\ve_\pp}f_{\mathrm{eq}} 
\end{align}
where the terms in square brackets reproduce the form of the intrinsic term, albeit with different frequency dependence and with $\Omega_d^{\rm sk} = \tau_\omega^2 \epsilon_{abd} \partial_a A^{\textrm{sk}}_b /2$ replacing the intrinsic Berry curvature. The last, round bracket is a term due to skew acceleration. Clearly, analogous intrinsic terms obtained by $A^{\textrm{sk}}_a \rightarrow v_a/\tau$ or similar would vanish. In turn, the second rank pseudotensor dual to $\Lambda_{abc}^{\textrm{sk}}$ is given by
\begin{align}\label{eq:lambdaskew}
&\lambda^{\textrm{sk}}_{dc}=\frac12e^3\tau\tau_\w^2\epsilon_{abd}\nonumber \\ 
&\times\int_{\pp}\left[\left(1+\frac{\tau_\w}{\tau}\right)
v_c\p_a A^{\textrm{sk}}_b  -
  \frac{\tau_\w}{\tau} A^{\textrm{sk}}_a\p_c v_b\right]\p_{\ve_\pp}f_{\mathrm{eq}}.
\end{align}
Using $\vv_\pp= \p_\pp\ve_\pp$, as well as $\vv_\pp \p_{\ve_\pp} f_{\mathrm{eq}}=\p_{\pp} f_{\mathrm{eq}}$, and integrating by parts, it is easy to show that this tensor is traceless.

\subsection{Side jump mechanism}
\label{sec:sidejump}

The intrinsic mechanism considered above stems from the interband coherences induced by external electric fields as they accelerate the charge carriers. It is well known that collisions with impurities, which can also be viewed qualitatively as a sort of acceleration by the impurity electric field, also lead to the creation of interband coherence, which manifest itself through the appearance of a coordinate shift in impurity scattering. In this Section we discuss the contribution of such coordinate shifts, commonly referred to as ``side jumps'', into the current-induced Hall-like response in metals. 

For weak centrosymmetric impurity potential, the coordinate shift associated with a side jump event  - a net displacement of the center of the scattering wave packet - is given by \cite{Belinicher1982,Sinitsyn2006} 
\begin{align}\label{eq:sidejump}
  \d\rr_{\pp\pp'}=i \bra u_{\pp}|\p_\pp u_{\pp}\ket-i \bra u_{\pp'}|\p_{\pp'} u_{\pp'}\ket\nonumber \\ 
  -(\p_\pp+\p_{\pp'})\textrm{arg}\bra u_{\pp}|u_{\pp'}\ket.
\end{align}
The existence of the side jump leads to two modifications of the standard semiclassical transport theory. First, there is an extra contribution to the electric current, given by
\begin{equation}\label{eq:sjcurrent}
  \jj^{\textrm{sj}}=e\int_{\pp\pp'}w^{S}_{\pp\pp'}\d \rr_{\pp\pp'} \delta\left(\e_\pp-\e_{\pp'}\right) f_{\pp'}.
\end{equation}
Second, the energy-conserving delta-function in the impurity collision integral must be modified to take into account the work done by the external electric field during the side jump event: 
\begin{equation}\label{eq:sjcollisionintegral}
  I\{f_{\bm{p}}\}=-\int_{\pp'} (w_{\pp'\pp}f_{\pp}-w_{\pp\pp'}f_{\pp'})\d(\ve_{\pp}-\ve_{\pp'}-e\EE\d\rr_{\pp\pp'}).
\end{equation}In what follows, we restrict our considerations to the linear order in the semiclassical corrections to classical kinetics, \textit{i.e.} to $O(\delta \bm{r})$ order. This means that in products of the coordinate shift of Eq.~\eqref{eq:sidejump} and $w_{\pp,\pp'}$ we can retain only the symmetric part of the latter - this amounts to neglecting the interplay between the skew scattering and side jump mechanisms. Further, seeking the $O(E^0 E^\w)$ contribution to the current, one needs to substitute the distribution function calculated to the $O(E^0 E^\w)$ order into Eq.~\eqref{eq:sjcurrent} for the side-jump accumulation current, as well as use the $O(E^0)$ and $O(E^\w)$ distribution functions in the collision integral~\eqref{eq:sjcollisionintegral}, after expanding it the linear order in $\d\rr_{\pp\pp'}$. 

To facilitate further progress, we introduce the side-jump accumulation velocity \cite{sjvelocitysign}: 
\begin{equation}\label{eq:sjvelocity}
  \vv^{\textrm{sj}}_{\pp}=\int_{\pp'}w^{S}_{\pp\pp'}\d \rr_{\pp ' \pp} \delta\left(\e_\pp-\e_{\pp'}\right).
\end{equation}
Then substituting the distribution function from Eq.~\eqref{eq:nonlineardistribution} into Eq.~\eqref{eq:sjcurrent}, and after an integration by parts, we obtain the accumulation part of the side jump current: 
\begin{equation}\label{eq:sjaccumulation}
  j^{\textrm{sj-\textrm{acc}}}_a=-e^3(\tau+\tau_\w)\tau_\w E^{\w}_{b}E^{0}_{c}\int_{\pp} v_c\p_bv^{\textrm{sj}}_a 
\p_{\ve_\pp} f_{\mathrm{eq}}.
\end{equation}

In order to obtain the part of the side jump current associated with the modification of the energy-conserving $\d$-function in the collision integral, Eq.~\eqref{eq:sjcollisionintegral}, we expand the $\d$-function to linear order in the electric field. Just like in the case of skew scattering, one has to solve the linear response problem first, as the corresponding solutions enter the generation terms in the nonlinear case. At the linear response level, one takes the distribution functions in the collision integral to be the equilibrium ones, and the collision integral expanded to linear order in the electric field becomes a generation term from the so-called anomalous distribution correction. The derivation proceeds in the standard way~\cite{SinitsynReview}, resulting in the following linear corrections to the distribution functions:
\begin{align}
\delta f^{\EE_0}_{\pp}=\tau e\EE^0\vv^{\textrm{sj}} \p_{\ve_\pp}f_{\mathrm{eq}},\quad
\delta f^{\EE_w}_{\pp}=\tau_\w e\EE^\w\vv^{\textrm{sj}}\p_{\ve_\pp}f_{\mathrm{eq}} 
\end{align}

To obtain the nonlinear corrections to the distribution function, one substitutes the perturbed distribution functions, Eqs.~\eqref{eq:linearizeddf}, into the collisions integral, and expands it to linear order in the electric field: no higher terms are necessary if we limit ourselves to corrections linear in the side-jump length. Retaining only the terms that oscillate at the optical frequency, we obtain the following expression for the nonlinear anomalous distribution correction: 
\begin{align}\label{eq:anomalousdistribution}
&\delta f^{\EE^0\EE^\w}_{\pp}=-\tau_\w e\EE^0\p_\pp \delta f^{\EE^\w}_{\pp}-\tau_\w e\EE^\w\p_\pp \delta f^{\EE_0}_{\pp} 
\nonumber \\ 
&+\tau_\w\int_{\pp'} w^S_{\pp\pp'}(f^{\EE^0}_{\pp}-f^{\EE^0}_{\pp'})e\EE^\w\d\rr_{\pp\pp'}\p_{\ve_\pp}\d(\ve_{\pp}-\ve_{\pp'})
\nonumber\\
&+\tau_\w\int_{\pp'} w^S_{\pp\pp'}(f^{\EE^\w}_{\pp}-f^{\EE^\w}_{\pp'})e\EE^0\d\rr_{\pp\pp'}\p_{\ve_\pp}\d(\ve_{\pp}-\ve_{\pp'}).
\end{align}
The corresponding part of the side jump current due to anomalous distribution is 
\begin{align}
\jj^{\textrm{sj-ad}}=e\int_{\pp} \vv_\pp \delta f^{\EE^0\EE^\w}_{\pp}. 
\end{align}
To make further process, we note that $\vv_\pp\p_{\ve_\pp}\d(\ve_{\pp}-\ve_{\pp'})=\p_\pp\d(\ve_{\pp}-\ve_{\pp'})$, and perform integration by parts to remove the derivative  from the $\d$-function. Finally, by noticing that the distribution functions in the collision terms vary faster than the scattering rate, we obtain the final form for the anomalous distribution current:
\begin{align}\label{eq:sjanomalousdistribution}
&j^{\textrm{sj-ad}}_a=e^3\tau \tau_\w E^{0}_{c}E^{\w}_{b}
\nonumber \\ 
&\int_{ \pp}\left(v_c\p_a v^{\textrm{sj}}_b+\frac{\tau_\w}{\tau}v^{\textrm{sj}}_b\p_c v_a+v^{\textrm{sj}}_c\p_b v_a+ \frac{\tau_\w}{\tau}v_b\p_a v^{\textrm{sj}}_c\right)\p_{\ve_\pp}f_{\mathrm{eq}}.
\end{align}
The total side-jump related current is obtained by combining the side jump accumulation, Eq.~\eqref{eq:sjaccumulation}, and anomalous distribution, Eq.~\eqref{eq:sjanomalousdistribution}, contributions. For the purpose of calculating the anti-symmetric part of the conductivity tensor, one can drop the last two terms on the right hand side of Eq.~\eqref{eq:sjanomalousdistribution}. Anti-symmetrizing the rest of contributions as appropriate, we obtain the final result for the side jump contribution to the GHE: 
\begin{align}
\Lambda^{\textrm{sj}}_{abc}=-e^3\tau\tau_\w\int_{ \pp}
\left[\left(1+\frac{\tau_\w}{2\tau}\right)
\left(v_c\p_b v^{\textrm{sj}}_a-v_c\p_a v^{\textrm{sj}}_b\right)\right.\nonumber \\ 
+\left.\frac{\tau_\w}{2\tau}\left(v^{\textrm{sj}}_a\p_c v_b-v^{\textrm{sj}}_b\p_c v_a\right)\right]\p_{\ve_\pp}f_{\mathrm{eq}}.
\end{align}
The corresponding dual tensor,
\begin{align}\label{eq:lambdasj}
&\lambda^{\textrm{sj}}_{dc}=e^3\tau\tau_\w\e_{abd} \nonumber \\ 
&\times\int_{\pp}
\left[\left(1+\frac{\tau_\w}{2\tau}\right)v_c\p_a v^{\textrm{sj}}_b-\frac{\tau_\w}{2\tau}v^{\textrm{sj}}_a\p_c v_b\right]\p_{\ve_\pp}f_{\mathrm{eq}},
\end{align}
is traceless. Note also that there is obvious similarity between the structure of Eqs.~\eqref{eq:lambdaskew} and~\eqref{eq:lambdasj}.

\subsection{Optical regime of high frequencies}
\label{sec:HighFreq}

In the high-frequency regime, the dominant contribution to the current-induced Hall-like optical response can be obtained from the usual expression for the optical conductivity, 
\begin{align}\label{eq:opticalsigma}
&\sigma_{ab}(\w)=ie^2\int_{\pp}\sum_{m\neq n}
\frac{v_{mn}^{a}v_{nm}^{b}-v_{mn}^{b} v_{nm}^{a}}{\ve^2_{nm}-(\w+i/\tau)^2}f_{n},
\end{align}
where $\vv_{nm}=\bra u_n|\p_\pp \hat h_\pp|u_m\ket$ are the matrix elements of the velocity operator and we assume the Fermi energy to reside in band $n$. The GHE is obtained by the straightforward replacement of the equilibrium distribution functions $f_n$ in band $n$ with those perturbed by the DC electric field: $f_n\to -\tau e\EE^0\p_\pp f_{\mathrm{eq}}$ \cite{note_taun}. The corresponding gyrotropic Hall response tensor is then determined by 
\begin{align}\label{eq:opticalLambda}
&\Lambda_{abc}(\w)=-ie^3 \tau \int_{\pp}\sum_{m\neq n}
\frac{v_{mn}^{a}v_{nm}^{b}-v_{mn}^{b} v_{nm}^{a}}{\ve^2_{nm}-(\w+i/\tau)^2}\p_c f_{\mathrm{eq}}(\ve_n).
\end{align}

For a generic two-band system, this expression can be further simplified by noting that 
\begin{equation}
    v_{mn}^{a}v_{nm}^{b}-v_{mn}^{b} v_{nm}^{a} = i \varepsilon_{nm}^2 \epsilon_{abc} \Omega_c^{\rm int}.
\end{equation}
In this case, Eq.~\eqref{eq:opticalLambda} takes the same form as the first term in Eq.~\eqref{eq:Lambdaabc} under the replacement 
\begin{equation}
    \boldsymbol{\Omega}^{\rm int} \rightarrow \frac{\varepsilon_{nm}^2}{\varepsilon_{nm}^2 + \tau_\w^{-2}} \boldsymbol{\Omega}^{\rm int}.
\end{equation}
This observation motivates the interpolating function associated with Eq.~\eqref{eq:BerryOmega}. Note that the inaccuracy associated with replacing $\boldsymbol{\Omega}^{\rm tot}$ by $\boldsymbol{\Omega}^{\rm tot}_{\omega}$ in all terms of Eq.~\eqref{eq:lambdatensor} is smaller than the accuracy of our calculations which are controlled by $(\omega,1/\tau) \ll \varepsilon_{n\bar n}$ ($1/\omega \tau \ll 1$) at small (large) frequencies. 

\section{Applications}\label{sec:applications}

In this section we discuss the GHE for two exemplary time reversal invariant models: a simple model of a 3D Weyl semimetal with broken inversion symmetry as well as a minimal Hamiltonian describing 2D transition metal dichalcogenide films under strain. More technical details on the calculation of the response are relegated to Appendix~\ref{app:Examples}.

\subsection{Chiral Weyl semimetal}

\begin{figure}
  \centering
  \includegraphics[width=3in]{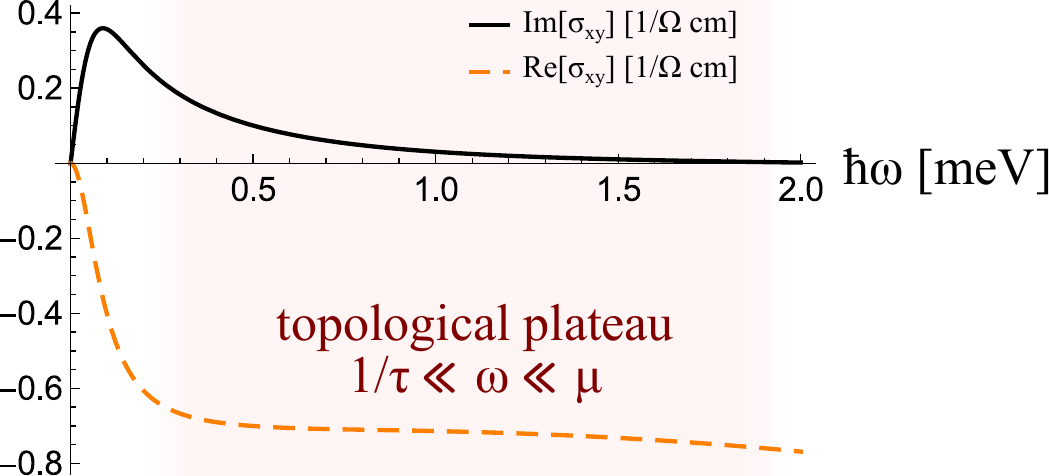}
  \caption{(Color online) The GHE in a minimal model of a time reversal symmetric Weyl semimetal according to Eq.~\eqref{eq:sigmaHall}. The plateau regime with substantial response is a measure of Berry monopoles. We used $\mu_+ = 10$ meV, $\mu_- = 3$ meV, $\hbar/\tau_+ = 1 meV$, {$\tau_- = \mu_+^2\tau_+/\mu_-^2$} and all other parameters as given in the main text, assuming $j_z = 10$ A/cm$^2$}\label{fig:WeylHall}
\end{figure}

\begin{figure}
  \centering
  \includegraphics[width=3in]{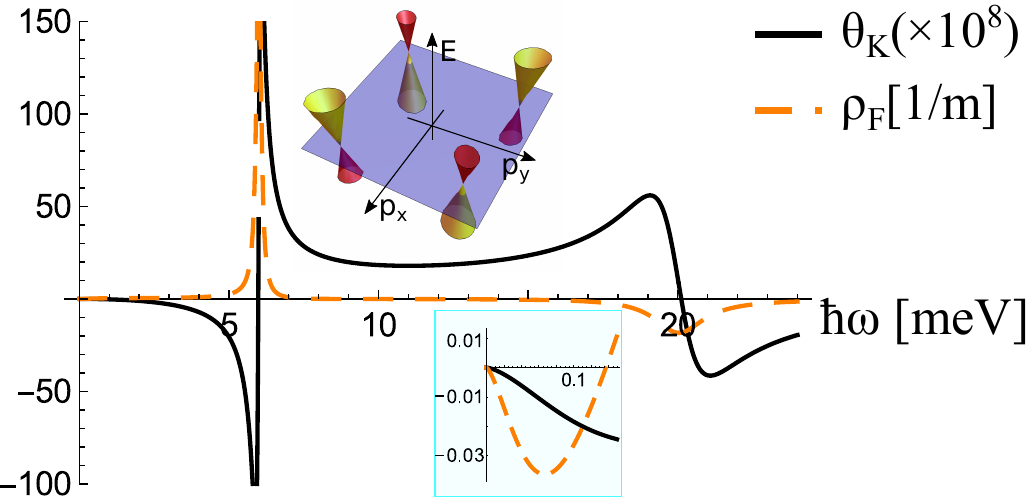}
  \caption{(Color online) Current induced optical activity in a minimal model of a time reversal symmetric Weyl semimetal. Main panel: Kerr angle and Faraday rotation per unit length for the same parameters as in Fig.~\ref{fig:WeylHall}. Upper inset: Dispersion of the model, the color code represents the sign of the Berry curvature in various bands. Lower inset: Low frequency regime. }\label{fig:Weylresponse}
\end{figure}

We consider a simple minimal model of a time reversal symmetric 3D Weyl semimetal,
\begin{align}
H &= \text{diag}\Big( v(\pp - \KK_+) \cdot \boldsymbol \sigma + E_+,v(\pp + \KK_+) \cdot \boldsymbol \sigma + E_+,\notag \\
&-v(\pp - \KK_-) \cdot \boldsymbol \sigma + E_-,-v(\pp + \KK_-) \cdot \boldsymbol \sigma + E_-\Big).\label{eq:WeylHamiltonian}
\end{align}
The band structure of this model contains two pairs of Weyl cones. Cones located at $\pm\KK_+$ are related by the time-reversal symmetry, thus having the same chirality. The same is true for the pair located at $\pm\KK_-$. Energy off-sets for each pair are given by $E_+\neq E_-$, which breaks inversion symmetry. The band structure is presented in Fig.~\ref{fig:Weylresponse} (upper inset). There is a substantial simplification in this model since each cone separately is perfectly isotropic, correspondingly side jump effects and skew scattering effects are vanishing.

The result for the gyrotropic Hall tensor is ($\hbar=1$)
\begin{equation}
\lambda_{ab} = \delta_{ab} \frac{e^3 }{6 \pi^2} [\Lambda(\mu_+,\tau_+) - \Lambda(\mu_-,\tau_-)]. \label{eq:Weylresponse}
\end{equation}
where $\mu_{\pm} = E_F - E_{+,-}$ have the meaning of the Fermi energy counted from the Weyl point for each pair of the cones, $\tau_+ \neq \tau_-$ are the transport mean free times, and 
\begin{equation}
  \Lambda(\mu,\tau) =\left(\tau-\tau_\w \right) \frac{4\mu^2}{4\mu^2 + \tau^{-2}_\omega}. \label{eq:WeylSingleNode}
\end{equation}

Using the Drude conductivity for this model, 
\begin{align}
    \sigma_D = \frac{e^2}{3 \pi^2 v } {(\mu_+^2\tau_+ + \mu_-^2\tau_-)},
\end{align}
we can express the GHE conductivity through the applied DC current, $j_z=\sigma_D E^0_z$: 

\begin{equation}
\sigma_{xy} = e j_z v \frac{\Lambda (\mu_+, \tau_+) - \Lambda(\mu_-, \tau_-)}{2 (\mu_+^2 \tau_+ + \mu_-^2 \tau_-)}. \label{eq:Weylconductivity}
\end{equation}

The typical values of parameters in the above equation are~\cite{DosReisHassinger2016,ArmitageVishwanath2018} $j_z = 10$ A/cm$^2$ and $v = 4 \cdot 10^7$ cm/s, which we use in Fig.~\ref{fig:WeylHall} to plot Eq.~\eqref{eq:sigmaHall}. In Fig.~\ref{fig:Weylresponse}, we illustrate the polar Kerr and Faraday rotation signals that stem from~\eqref{eq:Weylconductivity} and Eqs.~\eqref{eq:Kerr} and~\eqref{eq:Faraday}. 

We would like to emphasize here that the present considerations are restricted to noncentrosymmetric crystals in general, and Weyl systems in particular. They cannot explain the large magnitude of polarization rotation observed in Ref.~\cite{Xiu2015} in a centrosymmetric Dirac material Cd$_3$As$_2$.

\subsection{2D transition metal dichalcogenide}\label{sec:TMD}

\begin{figure}
  \centering
  \includegraphics[width=3in]{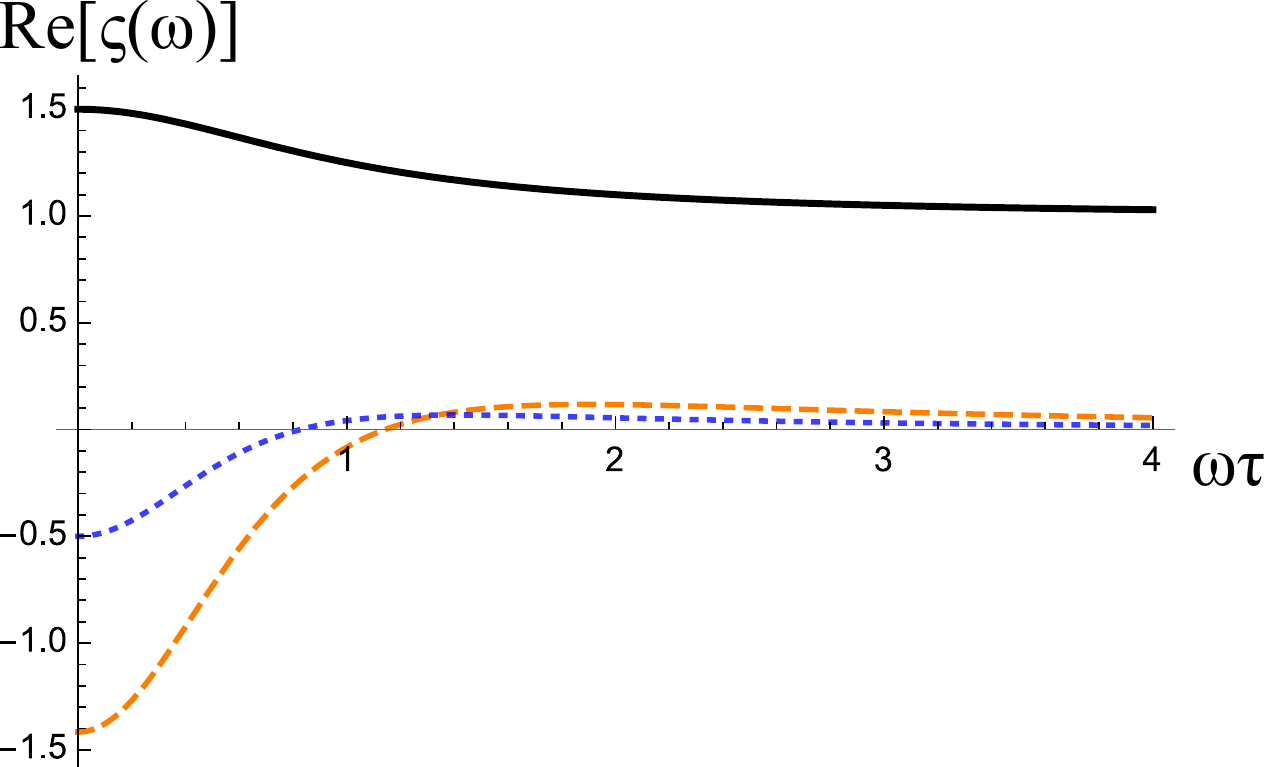}
  \includegraphics[width=3in]{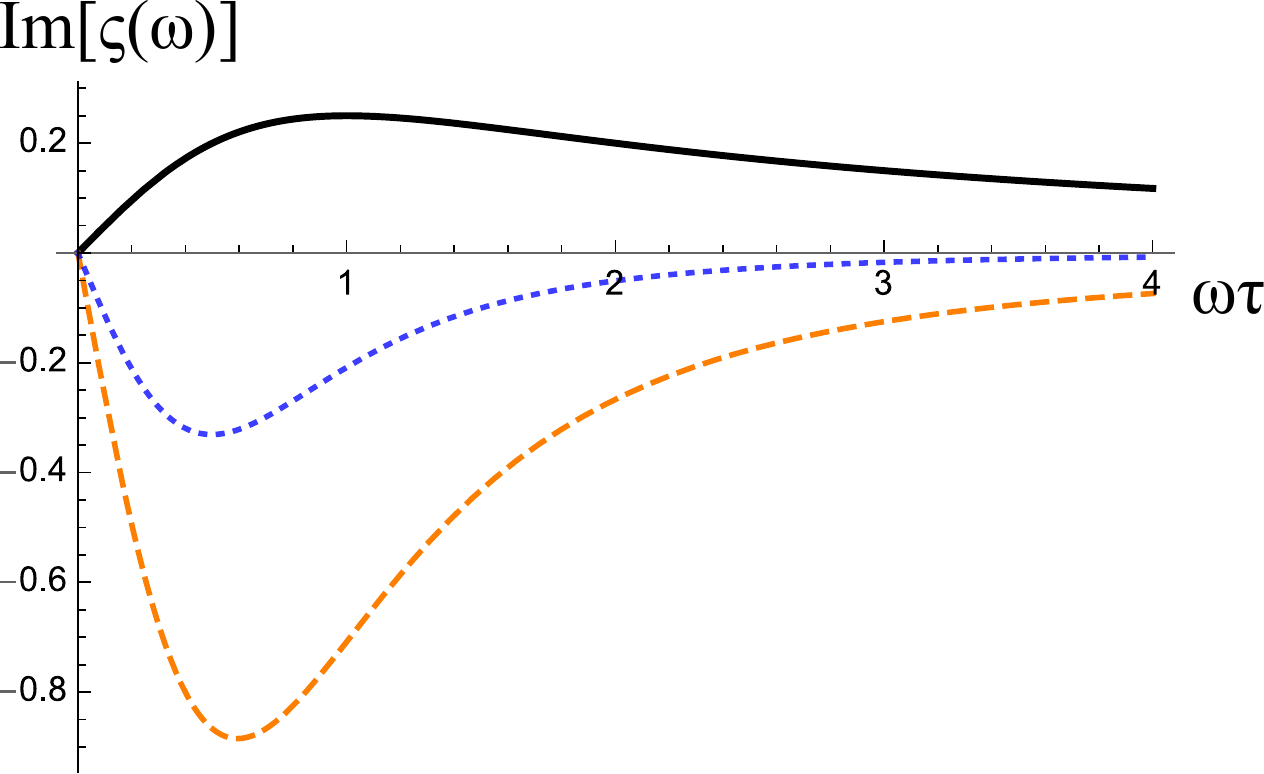}
   \caption{Frequency dependence of the real (upper panel) and imaginary (lower panel) parts of the gyrotropic Hall tensor for intrinsic (black solid line), skew-scattering (blue dotted line), and side-jump (orange dashed line) contributions for the 2D TMDs. To produce these plots we chose $m\tau^2/\tau_{\mathrm{sk}}=1$.}
   \label{fig:Lambda-vs-tau}
\end{figure}

In this section we consider the response of a strained 2D TMD as analyzed experimentally in Ref.~\cite{LeeShan2017} for MoS$_2$ monolayes. The Hamiltonian in valley $\xi = \pm 1$ takes the form 
\begin{equation}
H= v_x p_x \sigma_x + \xi v_y p_y \sigma_y + m \sigma_z + \xi v_y \beta p_y. \label{eq:TMDHamiltonian}
\end{equation}

In contrast to the perfectly isotropic Weyl model of the previous section, this model allows extrinsic processes. However, despite the seeming simplicity of this Hamiltonian, the microscopic calculation of extrinsic contributions to gyrotropic tensor is in fact very laborious. To simplify matters, we restrict our attention to the regime of weak strain, i.e. $v_x=v_y=v$ and thus focus on the leading order in $\beta$. Regarding the skew scattering we concentrate on the effect of impurity scattering beyond the Born approximation and disregard Gaussian and diffractive contributions (this is valid in the limit of strong sparse impurities). In addition, to be consistent with the assumption of constant relaxation time approximation, we explore the limit when chemical potential $E_F$ lies close to the bottom of the conduction band, so that elastic scattering time $\tau$ for this model is practically momentum independent. The final result in the low-frequency limit $\omega \ll m$ shows the importance of extrinsic processes 
\begin{align}\label{eq:Lambda-TMD}
&\Lambda_{abc}=\epsilon_{ab}\delta_{cy}\frac{3e^3}{8\pi}\left(\frac{\beta v\tau}{m}\right)\left(\frac{vp_F}{m}\right)^2\big[\varsigma^{\mathrm{int}}+\varsigma^{\mathrm{sj}}+\varsigma^{\mathrm{sk}}\big],
\end{align}
where the dimensionless functions $\varsigma(\omega)$ are given by 
\begin{equation}
\begin{split}
&\varsigma^{\mathrm{int}}= 1+\frac{\tau_\omega}{2\tau},\quad
\varsigma^{\mathrm{sj}}=-\frac{\tau_\omega}{6\tau} \left(1+\frac{15\tau_\omega}{2\tau}\right), \\ 
& \varsigma^{\mathrm{sk}}=
-\frac{m\tau_\omega^2}{3\tau_{\mathrm{sk}}}
\left(1+\frac{\tau_\omega}{2\tau}\right). 
\end{split}
\end{equation}

These expressions are valid for $vp_F = \sqrt{E_F^2 - m^2} \ll m$, whereas general formulas are derived in Appendix~\ref{app:TMD} for arbitrary relation between $vp_F$ and $m$. There the skew-scattering time $\tau_{\mathrm{sk}}$ is also defined in terms of parameters of the model [see Eq. \eqref{eq:App-Ask}]. It is apparent that at low frequencies, when $\omega\tau\ll1$, intrinsic and extrinsic contributions are of the same order, however, they have different asymptotic behavior at large frequencies $\omega\tau\gg1$, where $\varsigma^{\mathrm{int}}\propto 1$,
$\varsigma^{\mathrm{sj}}\propto 1/\omega$ whereas $\varsigma^{\mathrm{sk}}\propto 1/\omega^2$. 

In order to examine the role of skew scattering, we note that for a clean material with weak impurities one has $\tau/\tau_{\mathrm{sk}}\lesssim 1$ and $m\tau\gg1$. At the same time, the relative importance of the skew scattering contribution depends on the parameter $m\tau^2/\tau_{\mathrm{sk}}$ that can be smaller or greater than one. When this parameter is large there exists a parametrically wide range of frequencies, up to $\omega<\sqrt{m/\tau_{\mathrm{sk}}}$, where skew mechanism dominates. 

Furthermore, the skew scattering always dominates at small frequencies for the case of a clean 2DEG with screened Coulomb impurities, if the dielectric constant of the surrounding material is not too large. In this case, as can be shown to follow from Eq.~\eqref{eq:App-Ask}, $\tau/\tau_{sk}\sim1$. Given that in a clean material $m\tau\gg1$, we have $m\tau^2/\tau_{\mathrm{sk}}\gg1$ for the strength of the skew-scattering contribution, which makes it the dominant contribution at small frequencies. We plot the relative strengths of the intrinsic and extrinsic contributions to the GHE response function Eq. \eqref{eq:Lambda-TMD} in Fig. \eqref{fig:Lambda-vs-tau} to illustrate their frequency dependence and facilitate possible comparison with experiments.   

\begin{figure}
  \centering
  \includegraphics[width=3in]{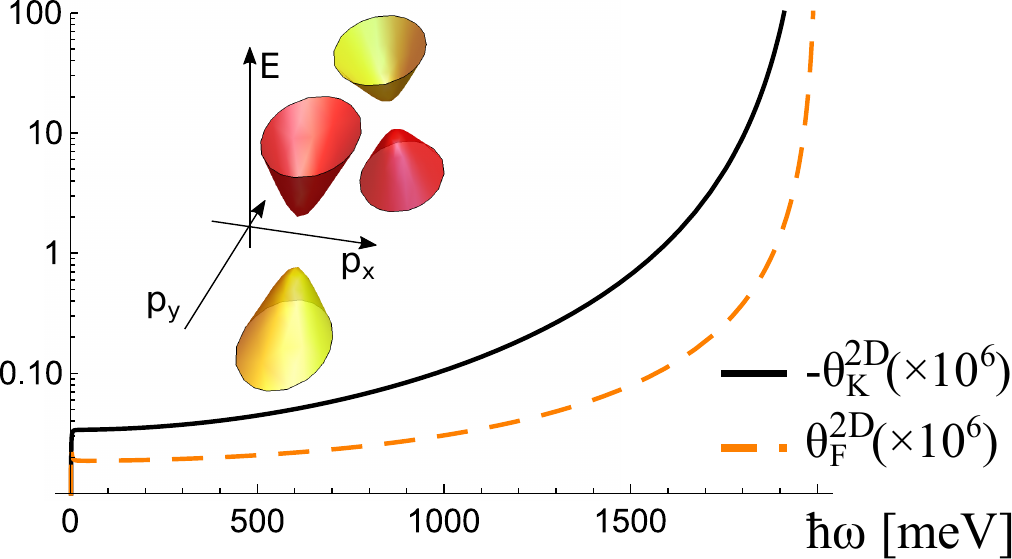}
  \caption{Current induced magneto optic effect of transition metal dichalcogenides. In this plot we used a typical current~\cite{LeeShan2017} $j_y = 10$ A/m, $m = 1eV$, $1/\tau = 1 meV$, $v = 10^7 cm/s$, $\sigma_{xx} = 50 e^2/h$, $\epsilon_0 c = 34 e^2/h$, $\beta = 0.1$, $m \tau^2/\tau_{\rm sk} = 1$.}\label{fig:TMDresponse}
\end{figure}

In Appendix~\ref{app:TMD} we also derive an interpolation formula for arbitrary $\omega$ which also captures the optical response $\omega \sim m$ of a TMD thin film described by Eq.~\eqref{eq:TMDHamiltonian}. As compared to Eq.~\eqref{eq:Lambda-TMD} it acquires additional factors which signal optical transitions. Readers interested in the functional dependence should consult the appendix, while here, using the formulae for Kerr and Faraday rotation presented in Sec.~\ref{sec:phenomenology}, we present a plot of the result, Fig.~\ref{fig:TMDresponse}.

\section{Summary, discussion and Outlook}\label{sec:discussion}

In summary, in this paper we have developed a comprehensive theory of the gyrotropic Hall effect, which at large frequencies determines the current induced optical activity. Our main result for the gyrotropic Hall tensor, Eq.~\eqref{eq:LambdaFullResult} and Fig.~\ref{fig:Cartoon}, is determined both by intrinsic and extrinsic effects whose microscopic origin is explicitly elucidated in Sec.~\ref{sec:derivations}. In contrast to older works which concentrated on the intrinsic effect at large frequency \cite{Vorobev1979} we were able to discuss the phenomenon of current induced optical activity in the entire spectrum from smallest all the way to optical frequencies. We thereby identify the range of frequencies much smaller than the band splitting (typically on the THz scale) to be most interesting: In this regime the trace of the gyrotropic Hall tensor $\lambda_{ab}$ displays topological quantization. At the same time, extrinsic, impurity induced (side jump and skew scattering) contributions are found to be comparable to the intrinsic effect. It worth mentioning that the Berry curvature dipole, which determines the intrinsic contribution to the GHE, and indirectly affects the side jump contribution, has been recently shown to get significantly enhanced near a topological transition in BiTeI~\cite{Sodemann2018}.

In order to illustrate the generic relevance of our theory we have presented its application to the current induced optical activity of two exemplary materials: A 3D noncentrosymmetric Weyl semimetal and a strained 2D transition metal dichalcogenide monolayer, see Figs.~\ref{fig:Weylresponse},\ref{fig:TMDresponse}. We also used the latter example to illustrate the possibility of disentangling various extrinsic effects by means of their frequency dependence, Fig.~\ref{fig:Lambda-vs-tau}.

As a major practical outcome, we suggest to use the low-frequency gyrotropic Hall effect as an efficient way to experimentally determine the topological nature of a 3D chiral metal. This proposal relies on the aforementioned observation that the trace of the gyrotropic Hall tensor of a given valley, Eq.~\eqref{eq:Trlambda}, is proportional to the valley's topological charge (the number of monopoles of Berry curvature enclosed in the Fermi surface). Therefore, it vanishes for non-topological materials while for topological systems, e.g. Weyl semimetals, the same quantity is substantial and, even more importantly, robust. 
The simplest experimental procedure to extract the trace (i.e. the spatial average) of $\lambda_{ab}$ is the analysis of the gyrotropic Hall effect in polycrystalline samples, see Fig.~\ref{fig:WeylHall}.

Another line of practical applications of magneto-optical phenomena related to the gyrotropic Hall effect can stem from the fact that its sign and amplitude are determined by the background current. Therefore, the GHE represents a nanoelectronic implementation of highly controllable optical activity, which previously was discussed only in metamaterials \cite{YuTian2016,ZhuTang2018} and correlated heterostructures \cite{HwangTokura2012}.

\section{Acknowledgements}

We acknowledge helpful discussions with Kin Fai Mak and Inti Sodemann. This work was supported by the National Science Foundation Grants No. DMR-1853048 (DAP), DMR-1653661 (AL), DMR-1506547 (MD) and the U.S. Department of Energy, Basic Energy Sciences, grants number DE-FG02-99ER45790 (EJK) and, in part, by DE-SC0016481 (MD).


 \appendix
\section{Application to 3D Weyl system and strained 2D TMDs}
 \label{app:Examples}

 In this section of the appendix we present details on the calculation of the response for the two exemplary models presented in the main text. We remind the reader of the notation $\xi = \pm 1$ for the helicity of a given node and $\zeta = \pm 1$ which labels the conduction and valence band index of a given two-band model. In this appendix we use the notation $\omega_+ = \omega + i /\tau$.

\subsection{General two-band model}

We briefly review some generic formulae for 2-band models of the form $H(\pp)=d_0(\bm{p})+\bm{d}(\pp) \cdot \bm{\sigma}$. From the wavefunctions $\vert{u_{\pp\zeta}}\ket = (d_3 + \zeta d, d_1 + i d_2)^T/\sqrt{2d(d+\zeta d_3)}$ it follows that 
\begin{eqnarray}\label{eq:uu}
&&\vert \braket{u_{\pp\zeta} \vert u_{\pp'\zeta}} \vert^2 = \frac{1 + \hat d \cdot \hat d'}{2} \\
&&\Omega_{a} = - \frac{\zeta}{4} \epsilon_{abc} \epsilon_{ijk} \hat d_i\partial_{b} \hat d_j \partial_c \hat d_k
\end{eqnarray}
where $\hat{d}_a=d_a/d$. We further investigate the amplitude for a closed path of three hoppings in momentum space
\begin{align}
    z_{\pp\pp'\pp''} &= \braket{u_\pp \vert u_{\pp'}} \braket{u_{\pp'} \vert u_{\pp''}} \braket{u_{\pp''} \vert u_{\pp}} \notag \\
    &= \frac{1 + (\hat d \cdot \hat d' + \hat d \cdot \hat d'' + \hat d' \cdot \hat d'') + i \zeta \hat d \cdot (\hat d' \times \hat d'')}{4}.
\end{align}
The imaginary part of $z_{\pp, \pp', \pp''}$ determines the skew scattering \cite{NagaosaReview}, while its phase (the Pancharatnam phase) determines the side jump displacement \cite{Sinitsyn2006}
\begin{align}
(\delta \bm{r}_{\pp\pp'})_a &= [\partial_{\pp''_a} \vert_{\pp'' \rightarrow \pp} + \partial_{\pp''_a} \vert_{\pp'' \rightarrow \pp'}] \arg\left(z_{\pp\pp' \pp''}\right) \notag \\
&= \zeta \frac{(\partial_a \hat d + \partial_a' \hat d') \cdot (\hat d \times \hat d')}{4\, \vert \braket{u_\pp \vert u_{\pp'}} \vert^2}. \label{eq:deltar}
\end{align}

 \subsection{Time reversal symmetric Weyl materials}

 For a single isotropic Weyl node the expressions for Berry curvature and velocity are (all momenta relative to the given Weyl node)
  \begin{eqnarray}
 \Omega_d = - \xi \zeta \frac{p_d}{2 p^3}, \quad
 v_c =\zeta v \frac{p_c}{p}.
 \end{eqnarray}
 
 Using the expression $\partial_{\epsilon_{\v p}} f_{\mathrm{eq}}(\epsilon_{\v p}) = - \delta(E_F - \zeta v p)$ we directly obtain the dc Drude conductivity and GHE response tensors per cone
 \begin{subequations}
 \begin{eqnarray}
 \sigma_D &=& \frac{e^2}{6 \pi^2 v}E_F^2 \tau \\
 \Lambda_{abc} &=& \xi \frac{e^3}{12\pi^2} \epsilon_{abc}  \frac{4E^2_F\left(\tau-\tau_\w \right)}{4E^2_F+ \tau^{-2}_\omega}.
 \end{eqnarray}
This is the origin of Eqs.~\eqref{eq:Weylresponse} and \eqref{eq:Weylconductivity} of the main text.
 \end{subequations}

 \subsection{Strained 2D transition metal dichalcogenides} \label{app:TMD}

In this Appendix we present details on the calculation of the GHE tensor for 2D transition metal dichalcogenides. We concentrate on the conduction band and expand to leading order in small strain effects. This allows us to set  $v_x = v_y =v$ and expansion to linear order in $\beta$. 
We will use
\begin{align}
  \ve_\pp= d + \xi\beta vp_y, \quad 
  \bm{d}(\bm{p})= (v p_x, \xi v p_y, m),
\end{align}
which directly implies the intrinsic Berry curvature
\begin{equation}
    \Omega^{\rm int} \equiv \xi \Omega = - \xi \frac{m v^2}{2 d^3}, 
\end{equation}
and the side jump displacement
\begin{equation}
\delta \rr _{\pp\pp'} = - \frac{\hat z \times(\pp' - \pp)}{2 \vert \braket{u_\pp \vert u_{\pp'}}\vert^2} \left [ \Omega^{\rm int} \frac{d}{d'} + (\Omega^{\rm int})' \frac{d'}{d} \right],
\end{equation}
where a prime on $\Omega_\xi$ or $d$ implies evaluation with $\pp'$. 

We assume an impurity potential $V(\bm{r})$ which is smooth on the scale of the lattice spacing but short ranged with respect to the Fermi wavelength. For impurities with sufficiently weak potential, the symmetric part of the transition probability is given by 
\begin{equation}
    w_{\pp\pp'}^{S} = 2\pi \vert \braket{u_\pp \vert u_{\pp'}} \vert^2 n_{\rm imp} V_0^2,
\end{equation}
where $V_0$ is the zero-momentum Fourier component of the disorder potential. Together with Eq. \eqref{eq:uu} this defines quantum and transport rates 
\begin{align}
    \frac{1}{\tau_q} &= \int_{\pp'} W^{S}_{\pp\pp'} = \pi n_{\rm imp}\nu(d)V^2_0{\left(1 + \frac{m^2}{d^2}\right)}, \\
    \frac{1}{\tau} &= \int_{\pp'} W^{S}_{\pp\pp'} (1 - \hat v \cdot \hat v')= \frac{d^2+3m^2}{2\tau_q(d^2+m^2)},
\end{align}
where density of states is $\nu(E) =\Theta(E-m) \vert E \vert/(2\pi v^2)$, which we computed to the leading order in $\beta$ and dropped terms which are linear in $\beta p_y$.

The contribution from the third order scattering to the asymmetric part of the scattering probability is \cite{NagaosaReview}
\begin{align}
    W_{\pp\pp'}^{A} &=- (2\pi)^2 n_{\rm imp} V_0^3 \delta(\ve_\pp - \ve_{\pp'})\notag \\
    &\int_{\pp''} \delta(\ve_\pp - \ve_{\pp''}) \mathrm{Im}[\braket{u_\pp \vert u_{\pp'}}\braket{u_{\pp'} \vert u_{\pp''}}\braket{u_{\pp''} \vert u_{\pp}}] \notag \\
    &= 4\pi^2 n_{\rm imp} \nu(d) V^3_0 \delta (\ve_\pp - \ve_{\pp'}) \notag \\
    &\Big(- \frac{m}{4 d} \hat z \cdot (\hat d \times \hat d') + \frac{d+d'}{d} \frac{\beta}{8} \hat y \cdot (\hat d \times\hat d')\Big). \label{eq:WA}
\end{align}
The linear in $\beta$ correction stems from the expansion of $\ve_{\pp'}$ in the delta function, the analogous correction due to $\ve_\pp$ drops out after partial integration.
The skew acceleration has thus the form 
\begin{subequations}
\begin{equation}\label{eq:App-Ask}
    \bm{A}^{\rm sk} = \frac{1}{\tau_{\rm sk}} \left \lbrace(\hat z \times \hat \pp) [\xi h_1 + \beta \hat p_y h_2] -  \hat x \beta h_3 \right \rbrace,
\end{equation}
where $\tau_{\rm sk}^{-1}=2\pi V_0  \sgn(m) \nu(m)/\tau$ determines the skewness in the distribution of the disorder potential, $\hat{\pp}=\pp/p$, $h_i = h_i(p)$ with
\begin{align}
    h_1 &={ v\frac{(d^2 - m^2)^{3/2}}{2d (d^2 + 3m^2)} ,} \\
    h_2 &= {v\frac{(d^4 - m^4)}{2 d^2 (d^2 + 3m ^2)},}\\
    h_3 &={ v\frac{(d^2 - m^2)}{2 (d^2 + 3m^2)}.}
\end{align}
\end{subequations}
Here, the first correction term (proportional to $\hat z \times \hat \pp$) stems from the expansion of the delta function in $W^{A}_{\pp\pp'}$, and the second ($\propto \hat x$) from the linear corrections in the round brackets of \eqref{eq:WA}. Corrections due to linear in $\beta$ contributions to the velocity vanish.

Similarly, the side jump velocity takes the form 
\begin{subequations}
\begin{equation}
     \bm{v}_{\pp}^{\rm sj} = (\hat z \times \hat p) [\xi g_1 + \beta \hat p_y g_2] -  \hat x \beta g_3,
\end{equation}
where
\begin{align}
    g_1 &={- \frac{\Omega p}{\tau} \frac{4 d^2}{d^2 + 3m^2}}\\
    g_2 &={\frac{\Omega}{v \tau} \frac{2(d^2 - m^2)d}{d^2 + 3m^2}},\\
    g_3 &={-\frac{\Omega}{v \tau} \frac{d(m^2 + 3d^2)}{d^2+3m^2 }}.
\end{align}
The first (second) linear correction stems from the correction to $\ve_\pp$ ($\ve_{\pp'}$) upon expansion of the delta function. 
\end{subequations}

The Berry curvature corrections due to skew scattering and side jump are
\begin{subequations}
 \begin{align}
     \Omega^{\rm sk} &= \frac{\tau_\omega^2}{2 \tau_{\rm sk}} \left [\frac{\xi h_1 + \beta \hat p_y h_2}{p} + \xi \partial_p h_1 + \beta \hat p_y \partial_p h_{2+3}\right] \notag \\
     \Omega^{\rm sj } &= \tau_{\omega} \left [\frac{\xi g_1 + \beta \hat p_y g_2}{p} + \xi \partial_p g_1 + \beta \hat p_y \partial_p g_{2+3} \right]
 \end{align}
\end{subequations}
where we used the shorthand notations $h_{2+3}=h_2+h_3$, $g_{2+3}=g_2+g_3$.
We can further expand the Berry curvature in $\beta$ to find 
\begin{equation}
\nonumber
\Omega^{\rm tot}=\xi \Omega_0^{\rm tot} + \beta \hat p_y \Omega_1^{\rm tot}.
\end{equation}

To obtain the response of the gyrotropic Hall effect we consider Eq.~\eqref{eq:LambdaFullResult} keeping in mind the antisymmetrization in $a \leftrightarrow b$. The first and second lines are expanded to leading order in $\beta$, note that the terms stemming from the distribution function and from the linear corrections to the velocity partly cancel out in the constant relaxation time approximation. This leads to $(vp = \sqrt{d^2-m^2})$
\begin{eqnarray}
    \Lambda_{abc}\vert_{1,2} &=& - e^3 \nu(d) v \beta \left (\tau + \frac{\tau_\omega}{2} \right) \epsilon_{ab} \delta_{cy} 
    \notag \\&\times &
    \frac{1}{2}\left (  \Omega_1^{\rm tot} \frac{vp}{d} - \frac{(vp)^2}{d} \frac{\partial}{\partial d} \Omega_0^{\rm tot}(d)\right).
\end{eqnarray}
The third line of Eq.~\eqref{eq:LambdaFullResult} leads to 
\begin{align}\label{eq:Lambda-TBM-3}
    &\Lambda_{abc}\vert_{3} = e^3 v^2 \beta \tau_\omega^2 \frac{\nu(d)}{d} \epsilon_{ab} \delta_{cy}  \Bigg (\frac{(vp)^2(g_{2}+g_{3})}{2 d^2} - g_3 - \frac{g_2}{2}
    \nonumber \\ 
    &+\frac{1}{2} \frac{\partial}{\partial d} \left \lbrace vp \left [1-\left(\frac{vp}{d}\right)^2 \right ] g_1\right \rbrace \Bigg) + \left\{g_j \rightarrow \frac{\tau_\omega}{2\tau_{\rm sk}}h_j\right\}.
\end{align}
The structure of these expressions is quite clear. For example, in the last expression the partial derivative term appears after expanding the Fermi distribution function up to the linear order in $\beta$ with the subsequent integration by parts, while the first and the last terms are the contributions from the side jump velocity and side-jump acceleration correspondingly. In total, the response per valley is 
\begin{align}\label{LabcFin1}
&\Lambda^{\mathrm{int+sj+sk}}_{abc}\vert_{1,2}=\epsilon_{ab}\delta_{cy}e^3v\tau \beta(m^2-d^2)\frac{\nu(d)}{d}\nonumber \\ 
&\times\left (1 + \frac{\tau_\omega}{2\tau} \right)
{\frac{1}{2}}\left [ \frac{\Omega_1^{\rm tot}}{\sqrt{d^2-m^2}} -  \frac{\partial \Omega_0^{\rm tot}}{\partial d} \right],
\end{align} 
which can be split respectively into intrinsic 
 \begin{align}
 \Lambda^{\mathrm{int}}_{abc}\vert_{1,2}
 =\Upsilon_{abc} \left [3\frac{m^2 - d^2}{2} \left ( 1+ \frac{\tau_\omega}{2\tau}\right)\right],
\end{align}
side jump term   
 \begin{align}
\Lambda^{\mathrm{sj}}_{abc}\vert_{1,2}=
 \Upsilon_{abc}\left [\left (1 + \frac{\tau_\omega}{2\tau} \right)
{\frac{\tau_\omega}{2\tau}} (m^2-d^2) \left (\frac{d^2-3 m^2}{d^2+3 m^2}\right.\right. \nonumber \\ 
\left. \left. -\frac{24 \sqrt{d^2-m^2} \left(d^6+9 d^4 m^2-d^2 m^4-9 m^6\right)}{d \left(d^2+3 m^2\right)^3}\right )\right], 
\end{align}
and skew scattering contribution  
\begin{align}
&\Lambda^{\mathrm{sk}}_{abc}\vert_{1,2}=\Upsilon_{abc} \Bigg [\left (1 + \frac{\tau_\omega}{2\tau} \right) \frac{\tau_\omega^2}{2 \tau_{\rm sk}}\frac{d^2 - m^2}{m}
\nonumber \\ 
&\times\Bigg (\frac{d^6+18 d^4 m^2+7 d^2 m^4+6 m^6}{2 \left(d^2+3 m^2\right)^2} \notag \\
&+\sqrt{d^2-m^2}\frac{d^8+24 d^6 m^2-78 d^4 m^4-48 d^2 m^6-27 m^8 }{2 d \left(d^2+3 m^2\right)^3}\Bigg) \Bigg]. 
\end{align}
The respective components corresponding to Eq. \eqref{eq:Lambda-TBM-3} are 
\begin{align}
&\Lambda^{\mathrm{sj}}_{abc}\vert_{3}= \Upsilon_{abc}  \left [\frac{\tau_\omega^2}{2\tau^2} \frac{{3d^6 + 23 d^4 m^2 + d^2 m^4 - 27 m^6}}{(d^2 + 3m^2)^2} \right ],
\notag\\
&\Lambda^{\mathrm{sk}}_{abc}\vert_{3}
= \Upsilon_{abc} \left [\frac{\tau_\omega^3}{4 \tau \tau_{\rm sk} m} \frac{(d^2-m^2)(d^4-m^4)(d^2+6m^2)}{(d^2+3m^2)^2} \right],
\end{align}
where we introduced $\Upsilon_{abc}=e^3\epsilon_{ab}\delta_{cy}(\beta v\tau)(\nu(d)\Omega/d^2)$. These terms can be grouped according to different mechanisms. In particular, concentrating on the conduction band, $m>0$, and expanding to the leading order in $d-m \simeq (vp)^2/2m$, using $\nu \Omega/d^2 \simeq -1/(4\pi m^3)$, one recovers Eq. \eqref{eq:Lambda-TMD} from the main text.

\subsection{Optical regime and plot of Kerr and Faraday rotation}

In the optical regime, the Berry curvature acquires a multiplication of $4d^2/(4d^2+\tau_\omega^{-2})$. To the level of accuracy of our calculation, it is sufficient to multiply the intrinsic contribution by a factor
\begin{equation}
\frac{4 d^2 \left(12 d^2+\tau \omega ^2\right)}{3 \left(4 d^2+\tau \omega ^2\right)^2},
    \label{eq:TMDOpticalFactor}
\end{equation}
and leave side jump and skew scattering contributions unchanged (they are suppressed at optical frequencies). Near the bottom of the conduction band this leads to 
\begin{equation}
    \sigma_{xy}(\omega) = \frac{3ej_y^{\mathrm{dc}} }{2}\frac{\beta v}{m^2} \left [\tilde \varsigma^{\rm int} + \varsigma^{\rm sj} +\varsigma^{\rm sk} \right].
\end{equation}
Here, 
\begin{equation}
    \tilde \varsigma^{\rm int} =\frac{4 m^2 \left(12 m^2+\tau^2_\omega\right)}{3 \left(4 m^2+\tau^2_\omega \right)^2} \varsigma^{\rm int}
\end{equation}
and we used the longitudinal conductivity
\begin{equation}
\sigma_{xx} ( \omega) = \frac{e^2}{2\pi} \frac{(v p)^2 \tau_\omega}{2d}.   
\end{equation}
 

\begin{thebibliography}{54}
\bibitem[{not({\natexlab{a}})}]{note_symmetric}
\bibinfo{note}{The fact that the first term in the square brackets of
  Eq.~\eqref{eq:nonlinskew} does not make a contribution to the antisymmetric
  part of the to the conductivity tensor has consequences for its frequency
  dependence. Hence, a question arises whether this property holds beyond the
  constant relaxation approximation. It is straightforward to show that this is
  indeed the case, and this is a property of the symmetric part of the
  collision integral.}
\bibitem[{sjv()}]{sjvelocitysign}
\bibinfo{note}{Note that the definition of the side jump velocity adopted in
  this work differs by a sign from that of Ref.~\cite{Rou2017}}.
\bibitem[{not({\natexlab{b}})}]{note_taun}
\bibinfo{note}{We remark that, to be precise, the relaxation rate $1/\tau$ in
  Eq.~\eqref{eq:opticalsigma} is the interband relaxation rate $1/\tau_{mn}$
  and in the expression of the non-equilibrium distribution function $\tau$ is
  the relaxation time $\tau_n$ of the band under consideration. This subtleties
  do not affect the overall structure of the optical response and were
  therefore dropped for notational convenience.}
\end{thebibliography}


\end{document}